\documentclass[fleqn,usenatbib]{mnras}

\usepackage{newtxtext,newtxmath}

\usepackage[T1]{fontenc}
\DeclareRobustCommand{\VAN}[3]{#2}
\let\VANthebibliography\thebibliography
\def\thebibliography{\DeclareRobustCommand{\VAN}[3]{##3}\VANthebibliography}

\usepackage{graphicx}	
\usepackage{amsmath}	
\usepackage{soul}
\usepackage{caption}
\usepackage{subcaption}
\usepackage{orcidlink}
\usepackage{multirow}
\usepackage{booktabs}	
\usepackage{placeins}
\usepackage[dvipsnames]{xcolor}
\usepackage{balance}
\usepackage{cancel}
\usepackage{siunitx}

\hypersetup{
    colorlinks = true,
    citecolor = {MidnightBlue},
    linkcolor = {BrickRed},
    urlcolor = {BrickRed}
}
\definecolor{editblue}{RGB}{227, 61, 144}

\newcommand{\code}[1]{\texttt{#1}}
\newcommand{\mbf}[1]{\mathbf{#1}}
\newcommand{\tup}[1]{\textup{#1}}
\newcommand{\bsys}[1]{\mbf{b}_\tup{sys}}
\newcommand{\afg}[1]{\mbf{a}_\tup{fg}}
\newcommand{\RNum}[1]{\uppercase\expandafter{\romannumeral #1\relax}}

\defcitealias{Kennedy2023}{K23}
\defcitealias{Burba2024}{B24}

\title[Bayesian power spectrum estimation with systematic effects]{Bayesian power spectrum estimation with modelling of systematic effects in delay-fringe rate space}

\author[Dutta et al.]{Sohini Dutta,$^{1}$\thanks{E-mail: sohini.dutta@manchester.ac.uk}\,\orcidlink{0000-0003-1140-1582}
Philip Bull,$^{1,2}$\,\orcidlink{0000-0001-5668-3101}
Jacob Burba,$^{1}$\,\orcidlink{0000-0002-8465-9341}
Michael J. Wilensky,$^{3,1}$\,\orcidlink{0000-0001-7716-9312}
Zheng Zhang,$^{1}$\,\orcidlink{0000-0002-9154-2803}
\newauthor Ainulnabilah Nasirudin$^{1}$ \orcidlink{0000-0003-2213-4547} 
\\
$^{1}$Jodrell Bank Centre for Astrophysics, University of Manchester, Manchester M13 9PL, UK\\
$^{2}$Department of Physics and Astronomy, University of Western Cape, Cape Town 7535, South Africa\\
$^{3}$Department of Physics and Trottier Space Institute, McGill University, 3600 University Street, Montreal, QC H3A 2T8, Canada
}

\date{Accepted XXX. Received YYY; in original form ZZZ}

\pubyear{\the\year{}}

\begin{document}
\label{firstpage}
\pagerange{\pageref{firstpage}--\pageref{lastpage}}
\maketitle

\begin{abstract}
Observing the Epoch of Reionisation using 21cm radio interferometry has proven to be a challenging task. Extraction of the extremely faint redshifted signal is complicated by the presence of bright foregrounds, radio frequency interference (RFI), and systematic artefacts. 
We discuss the challenge of accounting for systematic effects, particularly cable reflections, that appear in the visibility data obtained from 21cm interferometers. Cable reflections cause attenuated copies of the foreground signal to appear outside the `foreground wedge' region in which foreground contamination is supposed to be localised.
We build on the {\tt hydra-pspec} Gibbs sampler to implement a model of the systematics as a multiplicative effect in delay-fringe rate space. We include this model in the inference of the joint posterior distribution, in addition to the 21cm signal, its power spectrum, and foregrounds. This allows the systematics contribution to be marginalised, rather than filtering it out and causing additional signal loss. We demonstrate the method on simulated visibility data for a single baseline, showing that the 21cm delay power spectrum can be recovered well regardless of the location of the systematics in delay-fringe rate space. Our implementation is suitable for modelling other multiplicative factors on the visibilities, e.g. residual gain errors.
\end{abstract}




\section{Introduction}\label{sec:intro}

The Epoch of Reionisation (EoR) is a transitional period in the history of the Universe that remains largely unobserved in the radio regime. Rapid structure formation took place during this period, leading to the emission of X-ray and UV photons from luminous sources which in turn ionised the neutral hydrogen (\textsc{Hi}) permeating the intergalactic medium (IGM) \citep{pritchard2010, Loeb2013, koopmans2015cosmic}. Studying the progression of this ionisation process is expected to aid our understanding of both the large-scale development of the Universe and the small-scale astrophysics driving galaxy evolution \citep{Zaroubi2013}. 

The redshifted 21cm signal from the neutral hydrogen (\textsc{Hi}) in theory enables us to directly study this phenomenon by tracing the distribution of \textsc{Hi} in the IGM \citep{pritchard2010, Zaroubi2013}. One of the biggest challenges to this task is the presence of Galactic and extra-galactic foregrounds that are $\sim 10^4 - 10^5$ brighter than the redshifted 21cm signal at frequencies of order 100~MHz where the redshifted EoR signal is expected to arise. The radio interferometers used to observe the signal are also impacted by a range of systematic effects, which further complicate the analysis \citep{Kern2019}. 

Radio interferometers measure complex visibilities as a function of angular scale, frequency, and local sidereal time. It is often convenient to work in the Fourier spaces corresponding to each of these dimensions, which are the $uv$ (interferometer baseline) vector, delay, and fringe rate respectively. In the flat sky, quasi-static approximation, the first two of these can be mapped onto cosmological Fourier vectors $k_\perp$ (perpendicular to the line of sight) and $k_\parallel$ (parallel to it) through a simple rescaling \citep{Liu2020}. In these coordinates, bright foregrounds appear in a localised `wedge' region of the space due to their intrinsic spectral smoothness, which is modulated by the chromaticity of the interferometer baselines in a significant but limited way \citep{Datta2010, Chapman2014, Thyag2015}. The 21cm signal, which is not spectrally smooth, is present across the whole Fourier space however. Therefore, a `window' can be defined that is dominated by the 21cm EoR signal, and a `wedge' that is dominated by foregrounds.

This segregation of the signal from the contaminants enables a conservative `foreground avoidance' observing strategy. Rather than trying to model or subtract the foregrounds, we can simply ignore the Fourier modes in the wedge, focusing only on the EoR window to try and make a detection of the 21cm signal. This approach is complicated by instrumental effects that couple foregrounds into modes outside the wedge region however, such as cable reflections and mutual coupling \citep{Datta2010, Morales2012, parsons2012}.

In this work, we primarily model systematics that can be parametrised as a multiplicative factor (gain) to the sky signal, such as cable reflections, although many other forms of systematic effects exist. Impedance mismatches between the long transmission cables and other instrumental components introduce cable reflections in the data, producing reduced-amplitude copies of the signal at higher delay modes than the foreground-dominated modes. They typically peak over a few nanoseconds in the delay space, have the same time variation as the data, and contaminate the signal in the EoR window that is assumed to be otherwise free of foregrounds \citep{Kern2020a, Murphy2024}.

Another important systematic effect that affects data from interferometer arrays is mutual coupling, which arises from the electromagnetic interaction between interferometer elements. In this process, the electronic signal induced in one element by incident radiation is re-radiated and subsequently detected by a neighbouring element \citep{Kern2020a, Rath2024}. Like cable reflections, mutual couplings introduce additional spectral structures into the data and contaminate the EoR window. These appear as relatively localised artefacts in the delay-fringe rate space. In this space, the systematic artefacts are offset to non-zero delay modes that remain close to the foreground-dominated region. This is in contrast to cable reflections, which typically appear at higher delays further separated from the foreground dominated region. 

Although difficult to isolate in the frequency-time space, both of these systematics artefacts can be effectively visualised in the delay-fringe rate space \citep{Parsons2009, Kern2020a,Garsden2024}. Our multiplicative systematics model can represent any effect expressible as a gain term acting on a baseline's own sky signal, $\mathbf{d} = \mathbf{Y}\mathbf{s} + \mathbf{n}$, where $\mbf{d}$ is the data vector,  $\mbf{s}$ is the sky signal, $\mbf{Y}$ is the multiplicative systematic term, and $\mbf{n}$ is noise. Cable reflections satisfy this by construction because the resulting observed signal is the true signal plus a reduced-amplitude copy shifted to higher delay. Mutual coupling, however, does not reduce to a simple gain factor on the same baseline's own signal in the standard coupling model \citep{Kern2019}. Its leading-order contribution to a cross-correlation visibility is instead proportional to a different antenna's autocorrelation spectrum. Therefore, our method does not extend to this generalised case unless $\mathbf{Y}$ is permitted to act on multiple baselines (an extension that we leave to future work).

Motivated by this distinction, in this work we develop a method to model systematics present in the visibility data from an interferometer such as the Hydrogen Epoch of Reionization Array \citep[HERA; ][]{DeBoer2017}. HERA is custom built for the purpose of detecting the redshifted 21cm signal from the Cosmic Dawn and Epoch of Reionisation. It is situated in the Karoo desert in South Africa and consists of 350 dishes, each 14m in diameter. These dishes are densely packed in a (split) hexagonal grid. This maximises the number of redundant baselines, which are multiple antenna pairs with the same vector separation in both length and orientation. This increases HERA's sensitivity to low $k$ modes, i.e. the short baselines which are the most numerous in this layout. This is an important design optimisation, since the redshifted 21cm emission from the EoR is a diffuse background signal, with most of its power concentrated at large scales \citep{DeBoer2017}.

HERA uses the delay power spectrum (DPS) to measure the EoR spatial power spectrum. The DPS is formed by applying a Fourier transform along the frequency axis. It acts as a line-of-sight estimator that maps frequency fluctuations to delay modes \citep{parsons2012}. 

The reflection effects seen by HERA are also seen in other interferometers \citep{Jacobs2016, Beardsley2016, Mertens2025, Patil2016}. Cable reflections are a challenge particularly for the data observed in HERA Phase~I. In the Phase II~implementation of HERA, this issue was addressed by instrumental intervention. Longer cable lengths were used to place the reflections outside the range of delay modes of cosmological interest. Older coaxial cables were also replaced with fibre optic cables for their improved impedance matching properties \citep{Berkhout2024}. 

Several methods have been proposed to address the reflection systematics \citep{Kern2020a,Kern2020b, Murphy2024,Rath2024,Kern2025}. Some of these are based on fringe-rate filtering \citep{Parsons2014, Ali2015} to remove signals on particular temporal scales. However, aggressive filtering risks 21cm signal loss \citep{Garsden2024, Pascua2025}. Computationally expensive electromagnetic simulations can also be used to simulate cable reflections for an instrument, which can then be subtracted from visibility data \citep{Fagnoni2021}.

Hardware solutions like the introduction of extra coupling paths to destructively interfere with mutual coupling, improving impedance matching, feed placement, cable design and length considerations also aid in managing these systematic effects \citep{Fagnoni2021}. 

An innovative approach was proposed by \cite{Kern2019} that mathematically models systematics using temporal and spectral templates based on simulations. To model signal chain reflections, this method uses a non-linear parametrisation to calculate an initial estimate of the reflection parameters. To refine this estimate, they introduce perturbations into the initial guess, apply the calibration to the data in frequency space, transform this to Fourier space, and estimate the residual reflection bump. This is done iteratively until a minimum threshold of the reflection residuals is reached. The cross-couplings are modelled using a Singular Value Decomposition (SVD). An application of this method has already been demonstrated on HERA Phase~I data to achieve an order of magnitude improvement on the upper limits of the 21cm power spectrum predicted from this data set \citep{HERA2022}. \cite{Murphy2024} also proposed a method that uses a parametric forward model of the systematics, which is then constrained using a Hamiltonian Monte Carlo (HMC)-driven inference scheme optimised for modest numbers of highly non-linear instrumental systematics, with parameter-space gradients in the HMC improving sampling efficiency. 

Recently, \citet{Kennedy2023} and \citet{Burba2024} developed \texttt{hydra-pspec}, a Gibbs sampler that provides full Bayesian posterior inference and rigorous uncertainty quantification for the EoR signal and foreground components jointly. However, this framework does not currently model instrumental systematics that may be be present in the data, and their corresponding impact on the EoR/foreground signal uncertainties.

Although systematics removal methods have achieved considerable success, the impact of instrumental systematics on delay power spectrum estimation has not yet been fully characterised within a comprehensive uncertainty quantification framework, though it is an active area of development \citep{sims2016,sims2019a, sims2019b, burba2023, Kern2025}. This is especially important, as the systematics often overlap with the foregrounds and the EoR signal in the delay-fringe rate space, which can in principle lead to biases in the estimated power spectrum if not properly accounted for. Folding in systematic modelling within a Bayesian parameter estimation framework promises to provide a direct handle on these uncertainties.

We model the systematics as a multiplicative factor on the EoR and foreground signals, defined by the coefficients of a finite set of delay-fringe rate modes. We use a Gibbs sampler called  \code{hydra-pspec}\footnote{\url{https://github.com/HydraRadio/hydra-pspec}} \citep{Burba2024} to draw statistical samples from the joint posterior distribution of the parameters for all the components. Gibbs sampling is a Markov Chain Monte Carlo (MCMC) algorithm for drawing samples from a joint probability distribution over multiple variables when direct sampling is intractable, but the conditional distributions of each variable given all others are known and tractable \citep{Geman1984}. In the scenario presented here, the presence of thousands, even millions of parameters that need to be sampled to form the component posteriors makes Gibbs sampling a compelling choice.

We sample the separate components such as the EoR signal, foregrounds, and systematics via conditional distributions that are iterated over by the sampler. This forward-modelling and inference approach has several advantages beyond the rigorous propagation of uncertainties; successful predictions of missing or flagged data using such a method has been demonstrated in \cite{Kennedy2023} for example. \cite{Burba2024} additionally demonstrated the recovery of the delay power spectrum in the presence of uncertain foreground models using a similar method. These statistical recoveries have been found to be robust upon testing on simulated data \citep{Burba2024}.

The \code{hydra-pspec} code, first introduced in \cite{Kennedy2023} and later expanded in \cite{Burba2024}, implements a hybrid of Gaussian Constrained Realisations (GCR) and Gibbs sampling to perform delay power spectrum estimation. A GCR is simply a realisation of a Gaussian linear model with a Gaussian prior, conditioned on the data and other aspects of the data model. Realisations of the EoR signal parameters and foreground amplitudes are drawn jointly in the first step of the Gibbs sampler, followed by a sample of the EoR power spectrum, and then the systematics parameters. This combination forms a single Gibbs iteration. We repeat these steps thousands of times to sample the full joint posterior distribution of all the parameters.

The paper is organised as follows. We discuss the simulations and data model in Sect.~\ref{sec:methods}, including the multiplicative systematics model. In Sect.~\ref{sec:sampler}, we review the mathematical formalism and setup of the Gibbs Sampler. In Sect.~\ref{sec:results}, we present the results obtained from our experiments ranging over three test cases. This is followed by discussion and conclusions in Sect.~\ref{sec:conclusions}.
\section{Data generation and model setup}\label{sec:methods}

In this section we provide a detailed overview of the simulated visibility data we use in this work.
The simulations have been generated using \code{pyuvsim}, which is an extensively validated visibility simulator \citep{Aguirre2022,lanman2019pyuvsim,lanman2022validation}. The simulations themselves were previously used in \cite{Burba2024}, and include an EoR signal with realistic temporal structure (caused by the sky rotating through the primary beam of the instrument), and realistic foregrounds that have been modulated by the primary beam too.

We consider a reduced subset of the simulations for our analyses. The cropped data has 80 time samples and 60 frequency channels, reduced from the original 203 time samples and 120 frequency channels. This makes each test run less computationally expensive. The cropped visibilities have a frequency range of 100 -- 110~MHz, and correspond to a single baseline of length 14.6m. The choice of the 100--110 MHz frequency range was primarily motivated by the availability of simulated visibility data from \citet{Burba2024}, allowing us to reuse existing simulations while avoiding additional computational expense. This choice is not expected to affect the location of the simulated cable reflection features, which is determined by the cable length rather than the observing frequency. While the observing bandwidth influences the width of the reflection feature in delay space, its delay location remains unchanged for different frequency ranges of the same bandwidth \citep{Kern2020a}. Consequently, the selected frequency range is sufficient for investigating the impact of cable reflections considered in this work. Similarly, the reduction in the number of time samples does not affect our overall conclusions, since the cable reflection systematics we model are injected after this cut is applied. We also assume a Gaussian white noise level with an RMS that is $\sim 0.344\times$ that of the EoR field, i.e. we perform all our tests in the high signal-to-noise regime.

We follow a notation similar to \cite{Burba2024}, where lower-case bold letters (e.g. $\mbf{e}$) indicate vectors, and upper-case bold letters (e.g. $\mbf{E}$) indicate matrices. We begin by assuming that the signal $\mbf{s} = \mbf{e} + \mbf{f}$ is a linear combination of the EoR and foreground signals, where $\mbf{f}$ is a product of a set of spectral foreground basis vectors, encoded by the projection operator $\mbf{G}$, and some amplitudes (per mode and per LST) $\mbf{a}_\tup{fg}$, such that $\mbf{f}=\mbf{G}\mbf{a}_\tup{fg}$. The EoR signal $\mbf{e}$ is modelled as a Gaussian random field with mean zero and covariance matrix $\mbf{E} = \langle \mbf{ee}^{\dag}\rangle$, with angle brackets denoting an ensemble average. We note that this Gaussian random field assumption is an approximation. While the 21-cm signal is expected to become increasingly non-Gaussian as reionization progresses, particularly on smaller spatial scales, covariance-based Gaussian models provide a tractable description of the large-scale signal probed by current HERA observations and enable efficient Bayesian inference \citep{bharadwaj2005,mondal2015,majumdar2018,zhao2022,Aguirre2022}.

We now introduce a factor $\mbf{Y}$ that multiplies the signal $\mbf{s}$. This is intended to model the reflection systematics discussed above, but could incorporate other multiplicative contributions too, e.g. mis-estimated instrumental gains. The $\mbf{Y}$ operator is constructed from a set of delay-fringe rate modes and their amplitudes, which we will define shortly.

Note that modelling the systematics as a multiplicative term greatly reduces the number of parameters needed to accurately describe them; while it would also be mathematically simple to include systematics as an additive term, many more parameters would be needed to capture their structure, as (in the case of cable reflections) they are attenuated copies of the foregrounds, which are themselves relatively complex. In other words, an additive model essentially requires an entire copy of the signal to be modelled, whereas a multiplicative model requires only a few modes in the delay-fringe-rate space.

Under this model, the interferometer visibilities per time and frequency are
\begin{equation}
    V_\tup{ij}(\nu, t) = w(\nu,t)[Y_\tup{ij}(\nu,t) \left( e_\tup{ij}(\nu,t)+f_\tup{ij}(\nu,t) \right) +n_\tup{ij}(\nu,t)],
    \label{eq:vis}
\end{equation}
where $i$ and $j$ are the indices of a pair of antennas, $\nu$ and $t$ mark the observing frequency and time (LST) respectively, $\tup{Y}_\tup{ij}$ is the multiplicative systematic term, $n_\tup{ij}$ is the Gaussian noise, and $w$ is an element of a masking vector containing 0s and 1s, used to implement flagging (e.g. due to RFI). For every antenna pair or baseline, we can write the data model in a more compact form:
\begin{equation}
    \mbf{d} = \mbf{Y} \mbf{s} + \mbf{n},
    \label{eq:data_model}
\end{equation}
where $\mbf{d}$ is the data vector, $\mbf{Y}$ is the gain matrix representing the multiplicative systematic effect, $\mbf{s}$ is the signal (EoR plus foreground) vector, and $\mbf{n}$ is the noise vector.

To model the multiplicative systematic effect in delay-fringe rate space, only a few relevant Fourier modes are considered, rather than defining a complete Fourier basis. This is to reduce the number of free parameters; we would normally have a good idea of which delay-fringe rate modes the reflection structure occurs on, for example. The gain matrix $\mbf{Y}$ is formed such that its diagonal $\mbf{y}$ is defined as $\mbf{y}=\mbf{1}+\mbf{H}\mbf{b}_\tup{sys}$, where $\mbf{H}$ is the Fourier projection operator, populated only with selected modes, and $\mbf{b}_\tup{sys}$ is a vector of the (complex) amplitudes of these modes. In the absence of systematic effects, i.e. $\mbf{b}_\tup{sys} = \mbf{0}$, $\mbf{Y}$ reduces to the identity matrix.

For clarity, the shapes of all the vectors and matrices referred to from this section onwards are summarised in Table~\ref{tab:comp_shapes}. While some of these objects are quite large, it is often possible to avoid explicitly constructing them in the numerical implementation.
\begin{table*}
\centering
\resizebox{\textwidth}{!}{%
\begin{tabular}{llll}
\hline
Type & Component & Shape ($n_\tup{rows},n_\tup{columns}$) & Description \\
\hline

Matrix & $\mbf{Y}$   & $n_{\mathrm{\nu}} \times n_{\mathrm{t}} , n_{\mathrm{\nu}} \times n_{\mathrm{t}}$ & Systematic gain effect \\
       & $\mbf{N}$   & $n_{\mathrm{\nu}} \times n_{\mathrm{t}} , n_{\mathrm{\nu}} \times n_{\mathrm{t}}$ & Noise covariance matrix \\
       & $\mbf{E}$   & $n_{\mathrm{\nu}} , n_{\mathrm{\nu}}$ & Signal Covariance \\
       & $\mbf{G}$   & $n_{\mathrm{\nu}} , n_{\mathrm{FG}}$ & The foreground template\\
       & $\mbf{F}$   & $n_{\mathrm{FG}} , n_{\mathrm{FG}}$ & Foreground covariance\\
       & $\mbf{B}$   & $n_{\mathrm{sys}} , n_{\mathrm{sys}}$ & Systematic covariance \\
       & $\mbf{S}_\tup{g}$ & $n_{\mathrm{\nu}} \times n_{\mathrm{t}} , n_{\mathrm{\nu}} \times n_{\mathrm{t}}$ & Purely diagonal matrix with signal $\mbf{s}$ as the diagonal\\
       & $\mbf{H}$ & $n_{\mathrm{\nu}} \times n_{\mathrm{t}} , n_{\mathrm{sys}}$ & Fourier projection operator for the systematic term\\
\hline
Vector & $\mbf{e}$                & $n_{\mathrm{\nu}} \times n_{\mathrm{t}}$ & EoR signal\\
       & $\mbf{a}_{\mathrm{fg}}$  & $n_{\mathrm{t}} \times n_{\mathrm{FG}}$ & Foreground amplitudes\\
       & $\mbf{b}_{\mathrm{sys}}$ & $n_{\mathrm{sys}}$ & Systematic parameters \\
       & $\mbf{d}$                & $n_{\mathrm{\nu}} \times n_{\mathrm{t}}$ & Data vector\\
       & $\mbf{s}$                & $n_{\mathrm{\nu}} \times n_{\mathrm{t}}$ & Sky signal vector, sum of EoR and foregrounds \\
       & $\omega_\tup{n}$         & $n_{\mathrm{\nu}} \times n_{\mathrm{t}}$  & Random gaussian fluctuation vector wrt noise \\
       & $\omega_\tup{ns}$      & $n_{\mathrm{\nu}} \times n_{\mathrm{t}}$ & Random Gaussian fluctuation vector wrt square root of N\\
       & $\omega_\tup{e}$         & $n_{\mathrm{\nu}}$ & Random Gaussian fluctuation vector wrt EoR signal\\
\hline
\end{tabular}%
}

\caption{Summary of matrices and vectors used in the model. Here $n_{\nu},n_\tup{t},n_\tup{FG},n_\tup{sys},$ are numbers of frequency channels, times, foreground modes being modelled, and pairs of DL-FR modes being modelled for systematics. }
\label{tab:comp_shapes}
\end{table*}

\begin{figure*}
    \centering
    \includegraphics[width=\linewidth]{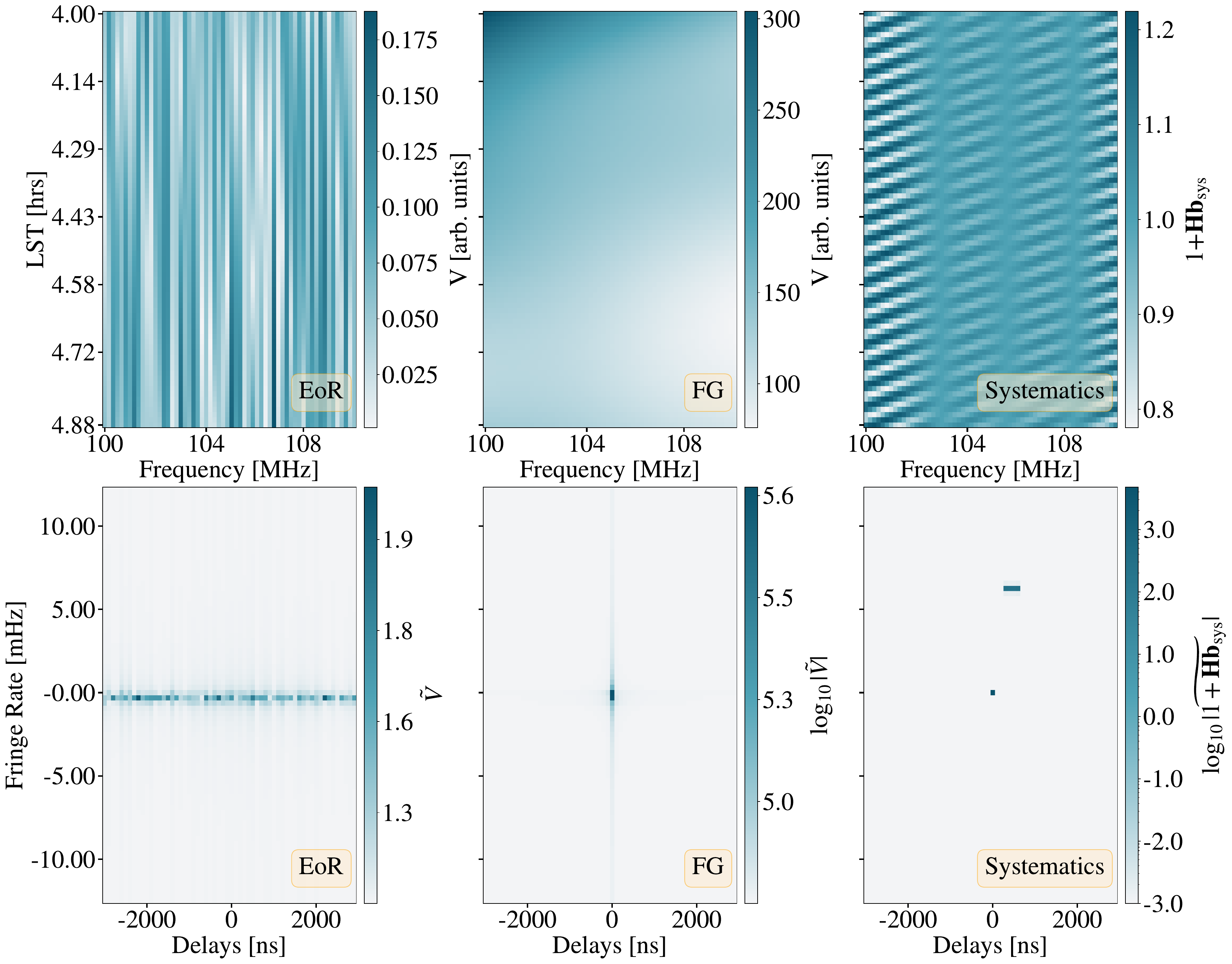}
    \caption[Absolute value of the example visibilities of the three data components in LST-frequency and delay-fringe rate space.]{The visibilities of the three data components, the EoR signal, the foregrounds and the systematics, represented in the LST-frequency space and the delay-fringe rate space. The top panel shows the quantities in the LST-frequency space and the bottom panel shows them in the delay-fringe-rate space. The EoR and foregrounds have been simulated using \texttt{pyuvsim} by \cite{Burba2024} and the systematic artefacts have been simulated by introducing contamination at four delay-fringe rate mode pairs. The panels here correspond to the systematics introduced in our third test case. The EoR and foreground visibilities in the top panel have been visualised in arbitrary visibility units, denoted by V and those of the bottom panel have been shown in the delay-fringe rate space, with $\tilde{V}$ denoting the Fourier transform of the same visibilities. The systematics panel in the top row shows the dimensionless quantity $1+\mathbf{H}\mathbf{b}_\mathrm{sys}$ and the systematics panel in the bottom row shows the same quantity in the delay-fringe rate space, with the Fourier transform of the same dimensionless quantity denoted by $\widetilde{1+\mathbf{Hb}}_\mathrm{sys}$. Note that in this figure, we have chosen not to apply a taper (as we have done in the subsequent figures). This results in some artefacts seen in the form of streaks at constant delays in the EoR and foreground figures in the bottom panel.}
    \label{fig:sys_in_dlfr}
\end{figure*}

\subsection{Modelling the systematic effects}
\label{sec:sysmodel}
We primarily concern ourselves with cable reflections in this work. A reflection in the signal chain of an antenna inserts a copy of the original signal with a time lag and an amplitude suppression. Following \cite{Kern2019}, this can be represented by
\begin{equation}
    v_1^\prime = v_1(\nu,t) + \epsilon_{11}(\nu)v_1(\nu,t),
\end{equation}
where $v_1^\prime$ is the voltage spectrum of antenna 1 with the reflection component and $\epsilon$ is a coupling coefficient describing the reflection in antenna 1's signal chain. The subscript denotes that the reflection is a result of the signal from antenna 1 coupling with itself. 

The reflections are localised to specific delay modes, depending on the length of the cable (or where the reflection occurs in the cable or a given component of the receiver). Mutual/cross-couplings can also be localised in this space. Because they are a copy of it, reflections appear at the same fringe rates as the sky signal. An example of a cable reflection feature can be seen in Fig. 11 of \cite{Kern2020b}. In Fig.~\ref{fig:sys_in_dlfr}, we show the simulated EoR, foregrounds, and systematics used in one of our test cases. All three are shown in frequency-time (upper panels) and delay-fringe rate (lower panels) spaces to highlight the localisation of the signals in the latter space.
\begin{figure*}
    \centering
    \includegraphics[width=0.9\linewidth]{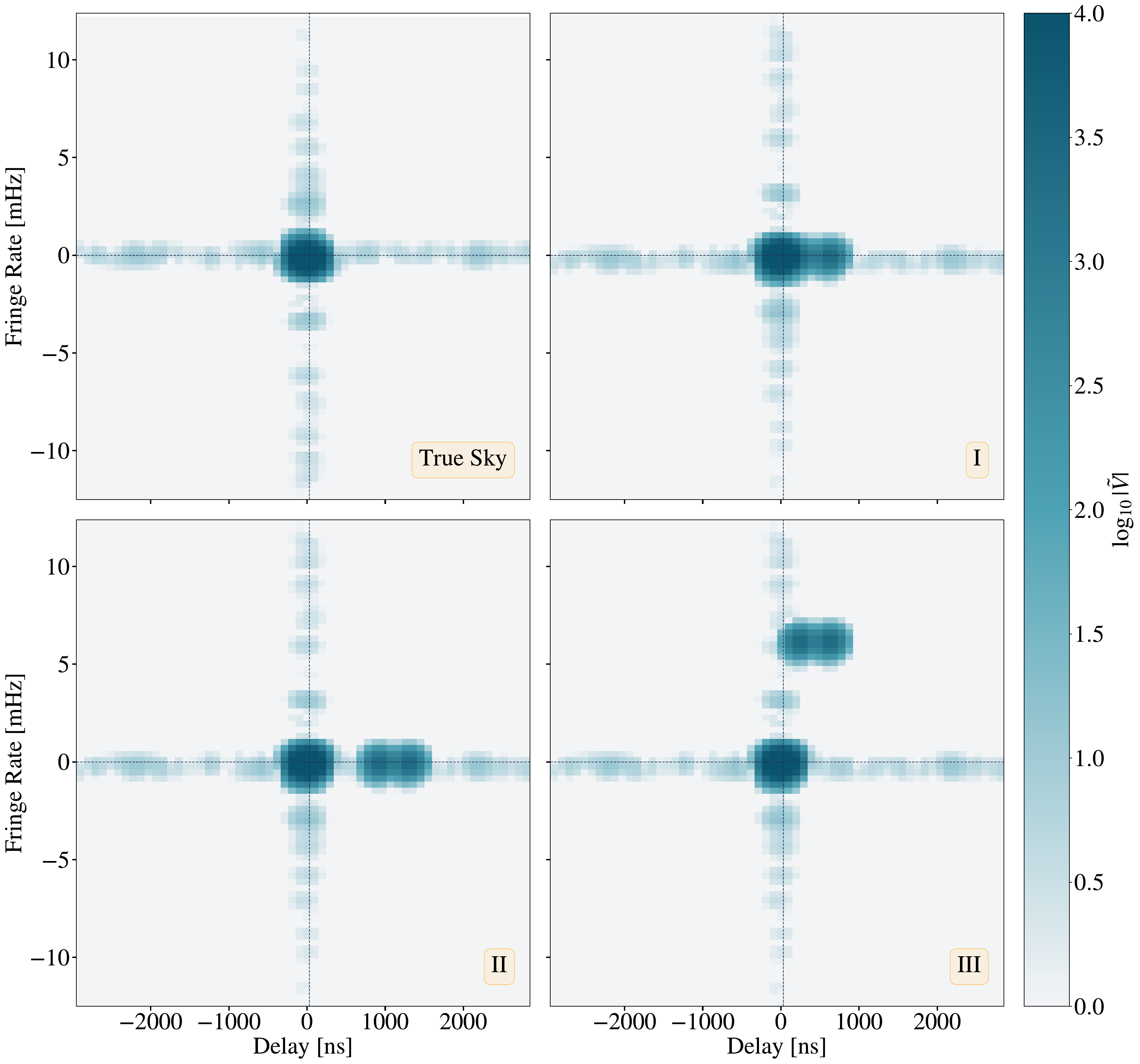}
    \caption{The three test cases we have considered in this work, shown in delay-fringe rate space. The first panel is the simulated data in the absence of any systematics, containing only the EoR and foreground visibilities (which we call the True Sky). Panels \RNum{1}, \RNum{2}, and \RNum{3} display the three different cases where the simulated systematics have been injected at different delay-fringe rate modes. In each of the cases, four delay-fringe rate mode pairs have been affected. We have applied the Blackman-Harris taper in this case before transforming to the delay-fringe rate space from the frequency-time domain.}
    \label{fig:test_data_dlfr}
\end{figure*}
The systematics models generated for the test cases studied in Sect.~\ref{subsec:res1} are formed by first selecting a handful of delay-fringe rate modes for each case, and then injecting artefacts that resemble a simplified version of cable reflections at those modes. We select four delay-fringe rate modes with an arbitrary set of parameter values, $\mbf{b}_\tup{sys,init} = (1 + 4i, 2 + 3i, 3 + 2i, 4 + 1i)$, forming a systematics factor $\mbf{Y}_\tup{init}$ by putting $\mbf{y}_\tup{init} = 1 + \mbf{H}\mbf{b}_\tup{sys,init}$ along the diagonal. The normalisation is such that, with these parameter values, $\delta g \sim 0.2$, i.e. a roughly 20\% effect on the uncontaminated visibilities. We use the same numerical values of $\mbf{b}_\tup{sys,init}$ for each case, despite the modes themselves being different, and form the simulated data using Eq.~\ref{eq:data_model}.

For concreteness, the modes being used are for the delay-fringe rate bins centred at $\tau = (295, 393, 492, 590)$~ns and $f = 0$~mHz for Case~\RNum {1}; the same delay modes but at $f = 6.25$~mHz for Case~\RNum{3}; and $\tau = (983, 1082, 1180, 1278)$~ns at $f = 0$~mHz for Case~\RNum{2}.
The resulting simulated data are shown in Fig.~\ref{fig:test_data_dlfr}. Despite only four modes being used for the systematics model, the features induced in the data are fuzzy `blobs' spread over several delay-fringe rate modes. There are two contributions to this spreading. One is that the discrete systematics modes modulate the foregrounds, which have power over a range of modes and so are mildly spread out in this space anyway. The second is that a window function (a Blackman-Harris taper) is applied to each dimension before transforming to delay-fringe rate space, in order to suppress ringing. This necessarily spreads out some of the signal too.

\section{The Gibbs Sampler}\label{sec:sampler}
To draw statistical samples from the posterior distribution for the three components with free parameters present in the data model for $\mbf{d}$, namely $\mbf{e}$, $\mbf{f}$, and $\mbf{Y}$, we use a Gibbs Sampler. A Gibbs Sampler is a Markov Chain Monte Carlo (MCMC) process that generates samples from a high-dimensional statistical distribution by iteratively sampling from a set of lower-dimensional conditional distributions. By iterating, this method can reconstruct the full joint posterior distribution, making it particularly useful when direct sampling is difficult but conditional distributions are tractable \citep{Geman1984}.

The joint posterior distribution we wish to sample from is
\begin{equation*}
    p(\mbf{e}, \mbf{a_\tup{fg}}, \mbf{b_\tup{sys}}, \mbf{E} | \mbf{G, N, H, B, d}),
    \label{eq:joint_post}
\end{equation*}
where $\mbf{N}$ is the noise covariance matrix defined by $\mbf{N}=\langle\mbf{nn}^{\dag}\rangle$, and $\mbf{B}$ is the systematics prior covariance. All other symbols retain their meanings from above. 

Using Bayes' theorem, we can write this as
\begin{align}
    &p(\mathbf{e, b_{sys}, a_{fg}}, \mbf{E} \mid \mathbf{G,N,H,B,d}) \nonumber \\ &~~~~~~~~~~~~\propto p(\mathbf{d} \mid \mathbf{e, b_{sys}, a_{fg}, H, G, N}) \, p(\mathbf{e, a_{fg}, b_{sys}}, \mbf{E} \mid \mathbf{B}).
    \label{eq:bayes}
\end{align}
The first factor on the right-hand side is the likelihood function, which is assumed Gaussian, and the second is the prior term. This system is further broken down into the conditional distributions
\begin{align}
    p(\mbf{s}|\mbf{b}_\tup{sys},\mbf{E},\mbf{H},\mbf{G},\mbf{N},\mbf{d}) &\propto p(\mbf{d}|\mbf{s},\mbf{b}_\tup{sys},\mbf{H},\mbf{N})p(\mbf{s}|\mbf{E},\mbf{G})
    \label{eq:condition1} \\
    p(\mbf{E}|\mbf{e}) = \frac{p(\mbf{e}|{E})}{p(\mbf{e})} &\propto \frac{1}{\tup{det}(\mbf{E})}\exp\left( -\mbf{e}^\dag \mbf{E}^{-1}\mbf{e}    \right)
    \label{eq:cov_eq} \\
    p(\mbf{b}_\tup{sys} | \mbf{s},\mbf{B},\mbf{N},\mbf{H},\mbf{E},\mbf{G},\mbf{d}) &\propto p(\mbf{d}|\mbf{b}_\tup{sys},\mbf{s},\mbf{H},\mbf{N})p(\mbf{b}_\tup{sys} | \mbf{B}),
    \label{eq:condition2}
\end{align}
where ${\rm det}()$ refers to a matrix determinant operation. Our notation indicates when the distributions are independent of particular parameters by omitting them from the conditioned-on part of the argument. We have also assumed that $\mathbf{e}$ is described by a complex Gaussian model with prior covariance $\mathbf{E}$, hence the specific Gaussian form of Eq.~\ref{eq:cov_eq}. 
These conditional distributions are sampled in the following Gibbs steps:
\begin{align}
    \mbf{e}_\tup{q+1}, \mbf{a}_\tup{fg,q+1} &\leftarrow p(\mbf{e},\mbf{a}_\tup{fg} | \mbf{b}_\tup{sys,q}, \mbf{H},\mbf{E}_\tup{q},\mbf{G},\mbf{N},\mbf{d})
    \label{eq:step_1a} \\
\mbf{E_\tup{q+1}} &\leftarrow p(\mbf{E} | \mbf{e}_\tup{q+1})    
\label{eq:step_1b} \\
    \mbf{b}_\tup{sys,q+1}  &\leftarrow p(\mbf{b}_\tup{sys} | \mbf{e}_\tup{q+1},\mbf{a}_\tup{fg,q+1},\mbf{B},\mbf{N},\mbf{H},\mbf{E}_\tup{q+1},\mbf{G},\mbf{d}),
    \label{eq:step_2}
\end{align}
where $\leftarrow$ indicates drawing a random sample from the distribution, and $q$ is the iteration index. By repeatedly cycling through this set of distributions and updating the free parameters at each iteration as indicated, we can eventually build up a set of samples that are drawn from the full joint posterior distribution.

\subsection{Gaussian constrained realisations}

Using the data model introduced in Eq.~\ref{eq:data_model}, and the set of conditional distributions given in Eqs.~\ref{eq:condition1}--\ref{eq:condition2}, we can build an iterative scheme to draw realisations of each component being sampled. Under the assumption of a Gaussian likelihood function, and either uniform or Gaussian priors on the signal and systematics coefficients, Eqs.~\ref{eq:condition1} and~\ref{eq:condition2} reduce to a set of linear equations that can then be solved for the parameters $\mbf{e}$, $\mbf{a}_\tup{fg}$ and $\mbf{b}_\tup{sys}$. This result is obtained by maximising the log-posterior of these conditional distributions to obtain the maximum a posteriori (MAP) solution (denoted by a hat). The MAP solution for the conditional distribution in Eq.~\ref{eq:condition1} is obtained by solving
\begin{align}
        \begin{pmatrix}
            \mbf{M}_{\rm ee} & \mbf{M}_{\rm ef} \\
            \mbf{M}_{\rm fe} & \mbf{M}_{\rm ff}
        \end{pmatrix}
        \begin{pmatrix}
            \mbf{\hat{e}} \\ \mbf{\hat{a}}_\tup{fg}
        \end{pmatrix}
        = 
        \begin{pmatrix}
            \mbf{Y}^{\dag}\mbf{N}^{-1}\mbf{d} \\
            \mbf{G}^{\dag}\mbf{Y}^{\dag}\mbf{N}^{-1}\mbf{d},
        \end{pmatrix}
    \label{eq:eor_fg}
\end{align}
where the matrix operator is
\begin{align}
    \begin{pmatrix}
            \mbf{M}_{\rm ee} & \mbf{M}_{\rm ef} \\
            \mbf{M}_{\rm fe} & \mbf{M}_{\rm ff}
        \end{pmatrix}
        \equiv 
    \begin{pmatrix}
           \mathbf{Y}^{\dag}(\mathbf{N}^{-1} + \mbf{E}^{-1})\mathbf{Y} & \mathbf{Y}^{\dag}\mathbf{N}^{-1}\mathbf{Y}\mbf{G} \\
            \mbf{G}^{\dag} \mathbf{Y}^{\dag}\mathbf{N}^{-1}\mathbf{Y} & \mbf{G}^{\dag} \mathbf{Y}^{\dag}\mathbf{N}^{-1}\mathbf{Y} \mbf{G} + \mbf{Y}^{\dag}\mbf{F}^{-1}\mbf{Y}
        \end{pmatrix}. \nonumber
\end{align}
This is similar to Eq.~9 in \cite{Burba2024}.
Similarly, the MAP solution for the distribution shown in Eq.~\ref{eq:condition2} can be written as
\begin{equation}
    \begin{pmatrix} \mbf{B^{-1}} + \mbf{H^{\dag} \mbf{S}_\tup{g}^{\dag} N^{-1} \mbf{S}_\tup{g} H} \end{pmatrix} \, \mbf{\hat{b}}_\tup{sys} =  \mbf{H^{\dag} \mbf{S}_\tup{g}^{\dag} N^{-1} d},
    \label{eq:Sys_solver}
\end{equation}
where $\mbf{S}_\tup{g}$ is a purely diagonal matrix with $\mbf{s}$ as the diagonal. This equation now has the form of a typical Wiener filter.

Next, we introduce fluctuation terms so that we can draw samples rather than simply solving for the MAP solution each time. The fluctuations are denoted by unit Gaussian random vectors of appropriate size, e.g. $\mathbf{\omega}_\textup{e}$ for the EoR field, and $\mathbf{\omega}_\textup{n}$ for the noise on the data vector in Eq.~\ref{eq:eor_fg}. Similar terms are also added to Eq.~\ref{eq:Sys_solver}, and they are scaled by inverse square roots of their corresponding covariances to produce Gaussian random draws with the correct covariance. We assume a very broad prior on the foreground model, and so the fluctuation term approximates to zero. The linear equations given in Eqs.~\ref{eq:eor_fg} and~\ref{eq:Sys_solver} are thus modified to obtain the `Gaussian constrained realisation (GCR)' solutions of the EoR signal and foreground amplitudes,
\begin{align}
        \begin{pmatrix}
            \mbf{M}_{\rm ee} & \mbf{M}_{\rm ef} \\
            \mbf{M}_{\rm fe} & \mbf{M}_{\rm ff}
        \end{pmatrix}
        \begin{pmatrix}
            \mbf{e} \\ \mbf{a}_\tup{fg}
        \end{pmatrix} 
        = 
        \begin{pmatrix}
            \mbf{Y}^{\dag}\mbf{N}^{-1}\mbf{d} + \mbf{E}^{-\frac{1}{2}}\omega_e + \mbf{Y}^{\dag}\mbf{N}^{-\frac{1}{2}}\omega_n \\
            \mbf{G}^{\dag}\mbf{Y}^{\dag}\mbf{N}^{-1}\mbf{d} + \mbf{G}^{\dag}\mbf{Y}^{\dag}\mbf{N}^{-\frac{1}{2}}\omega_n
        \end{pmatrix},
    \label{eq:GCR_eor_fg}
\end{align}
and systematics parameters
\begin{equation}
  \begin{pmatrix} \mbf{B^{-1}} \, + \, \mbf{H^{\dag} \mbf{S}_\tup{g}^{\dag}N^{-1}\mbf{S}_\tup{g}H} \, \end{pmatrix} \mbf{b}_\tup{sys} = \mbf{H^{\dag}\mbf{S}_\tup{g}^{\dag}N^{-1}d}+ \mbf{H}^{\dag}\mbf{S}_\tup{g}^{\dag}\mbf{N}^{-\frac{1}{2}}\omega_\tup{ns},
\label{eq:GCR_sys}
\end{equation}
where the random normal vectors $\omega_\textup{n}$, $\omega_\textup{ns}$ and $\omega_\textup{e}$ have zero mean and unit variance. The systematics covariance matrix $\mbf{B}$ is taken to be purely diagonal, with identical variances along the diagonal such that $\mathbf{B} = 100^2\,\mathbf{I}$. This is intended as a very broad prior. We provide a Gaussian prior on the systematics parameters with zero mean and obtain samples of the three components by solving Eqs.~\ref{eq:GCR_eor_fg} and~\ref{eq:GCR_sys} iteratively, updating the relevant parameter values after each random draw.

\begin{figure*}
    \centering
    \includegraphics[width=\linewidth]{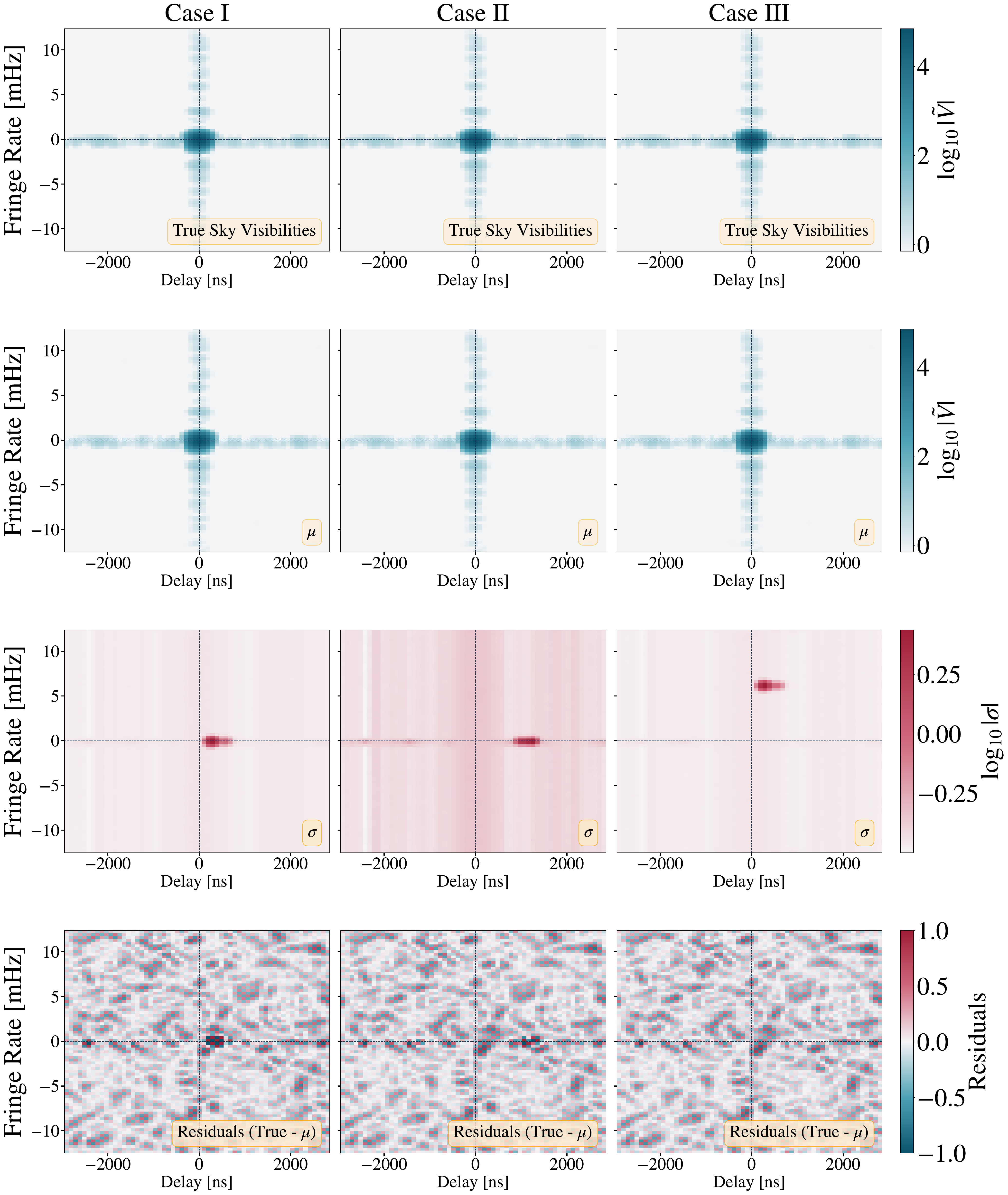}
    \caption{The recovered sky visibilities are compared against the true sky visibilities in this figure, visualised in delay fringe rate space. \textit{Top row}: The true sky visibilities, containing EoR and foregrounds for Cases \RNum{1} to \RNum{3} plotted from left to right. \textit{Second row}: The sky visibilities recovered from the sampler for each case. These were constructed by adding the EoR and foreground samples for each Gibbs iteration to form the sky visibilities and averaging over them to form the mean sky visibilities. \textit{Third row}: The standard deviations of the samples of sky visibilities. We can clearly see the imprint of the systematics artefacts here in the form of localised increase in the magnitudes of the standard deviations. \textit{Bottom row}: The residuals between the true sky visibilities and the mean of the samples for each case.}
    \label{fig:result_waterfalls}
\end{figure*}

\begin{figure*}
    \centering
    \includegraphics[width=\linewidth]{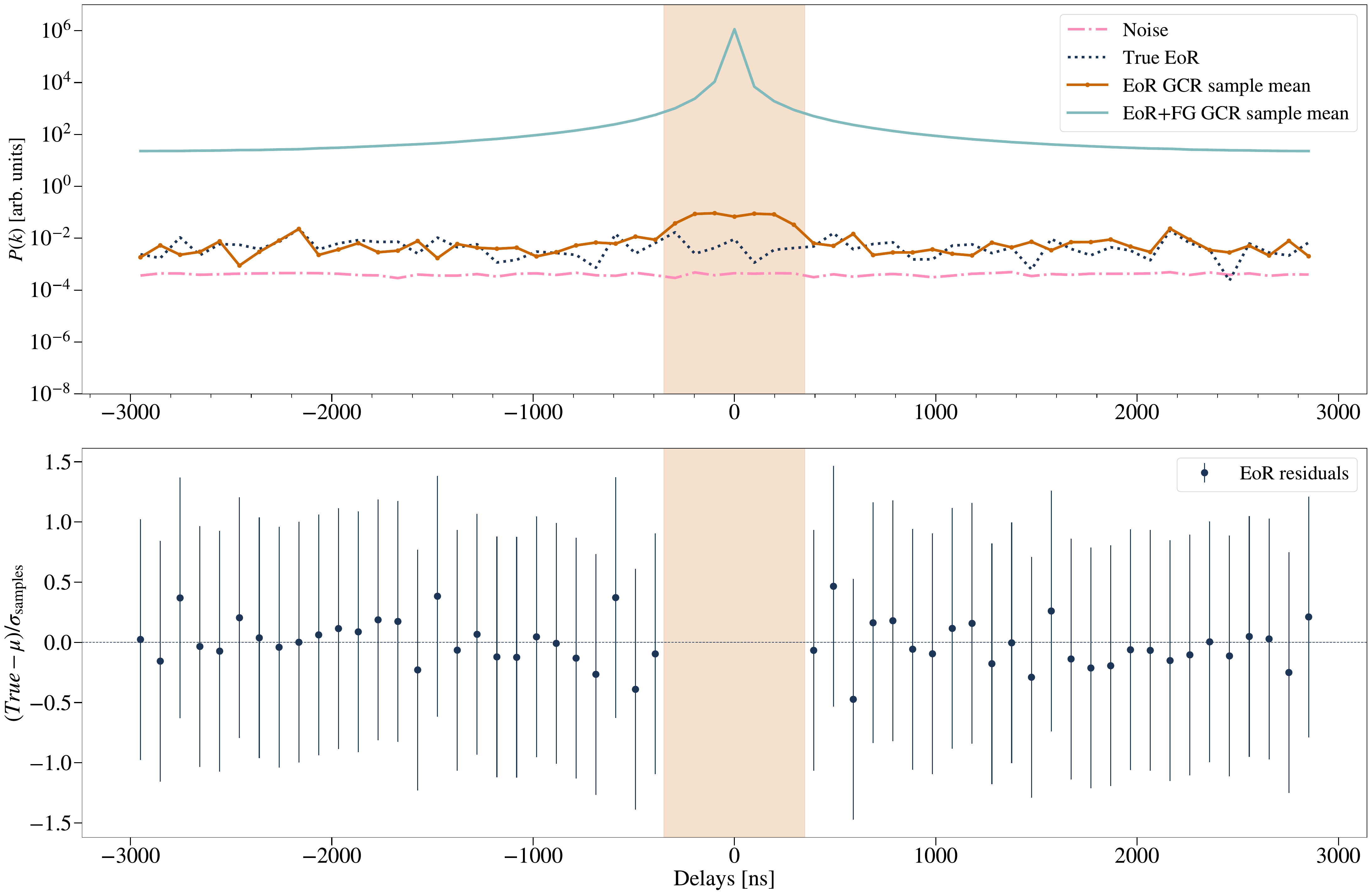}
    \caption{(Upper panel): The DPS of the recovered EoR and the foreground components, for Case~\RNum{3}. (Lower panel): Difference between the true and recovered EoR DPS, scaled by the standard deviation.}
    \label{fig:comp_errors}
\end{figure*}

\subsection{Sampling the EoR power spectrum}
\label{sec:eorpower}

The EoR signal covariance matrix $\mbf{E}$ does not have a Gaussian conditional distribution, as shown by Eq.~\ref{eq:cov_eq}. In general, this is an inverse Wishart distribution \citep{wishart2012}. By assuming statistical homogeneity and isotropy of the EoR signal as in \cite{Burba2024}, and working in delay space, the EoR covariance matrix becomes diagonal. We can then further simplify the conditional distribution into a product of independent inverse Gamma distributions for each EoR power spectrum band-power \citep{2008ApJ...676...10E, Kennedy2023}, conditioned on the EoR signal $\mbf{e}$ sampled from the first Gibbs Step. These distributions can be sampled from efficiently using inverse transform sampling in the second Gibbs Step given in Eq.~\ref{eq:step_1b}. This is done within pre-defined hard prior bounds, which are chosen as $10^{-7} < P(\tau) < 10^{-1}$ in the arbitrary delay power spectrum units used throughout the rest of this paper. These bounds were selected based on the delay power spectrum of the simulated EoR signal. They are chosen to comfortably encompass the expected range of true EoR delay power spectrum amplitudes, remaining sufficiently broad to avoid artificially constraining the parameter estimation, while still being narrow enough to prevent the sampler from exploring unrealistically small band-power values that would exacerbate the degeneracy between the foreground and low-delay signal modes discussed above. The weighting is uniform. This yields samples from a {\it truncated} inverse gamma distribution with a uniform prior.

We define the EoR covariance in delay space as $\mbf{\tilde{E}} = \mbf{T}^{\dag}\mbf{ET}$, where $\mbf{T}$ is the discrete Fourier projection operator from frequency ($\nu$) to delay ($\tau$) space. Thus, each component of the diagonal in $\mbf{\tilde{E}}$ represents the band-power $P(\tau)$ associated with the corresponding delay bin. The corresponding probability density function is
\begin{equation}
    f(x; \alpha, \beta) = \frac{\beta^{\alpha}}{\Gamma(\alpha)} x^{-\alpha-1} e^{-\beta/x},
    \label{eq:inv_gamma}
\end{equation}
where $\alpha$ is the shape parameter and $\beta$ is the scale parameter. The value of $\beta_{\tau}$, for each delay bin $\tau$, is computed from the Fourier-space EoR signal amplitudes, $\tilde{\mbf{e}} = \mbf{Te}$, for each Gibbs iteration $q$,
\begin{equation}
    \beta_{\tau.\tup{q}} = \sum_{t}^{N_t} \tilde{e}^{*}_{\tau,t,q} \, \tilde{e}_{\tau,t,q},
    \label{eq:beta}
\end{equation}
where $\tau$ is the central point of each delay bin, $t$ indexes the LST values, and $q$ is the Gibbs iteration index. The inverse Gamma distribution is a conjugate distribution for the variance of a Gaussian random field and follows directly from the inverse Wishart distribution for the full covariance matrix once the delay-space covariance is assumed to be diagonal. The shape parameter $\alpha$ for each inverse gamma distribution is the number of times (or `observations') minus 1, i.e. $\alpha = N_t - 1$, where we have assumed that each LST bin provides an independent observation of the field. (This is not actually the case however; as we will discuss later, there is a correlation length set by the beam crossing time.)

The inverse transform sampler works by mapping a probability value in the uniform distribution $u \in [0,1]$ to the corresponding value $x$ such that ${\rm CDF}(x) = u$ where ${\rm CDF}$ is the cumulative distribution function (CDF) of the inverse gamma distribution, multiplied by the prior density, between the pre-defined prior bounds. We provide the sampler with a uniform prior on the EoR delay power spectrum and obtain the sample $x$ by finding $x = {\rm CDF}^{-1}(u)$. 

Our treatment deviates slightly from \cite{Burba2024} in that the inverse gamma distributions are truncated, preventing unrealistic extreme values of the band powers from being drawn. While a prior range of $[0, \infty]$ for the band powers is statistically permissible, it can cause unwanted behaviour in the sampler. This occurs when a very small (but valid) band power value is drawn by the inverse Gamma sampler. This effectively imposes a very narrow prior on Eq.~\ref{eq:condition1} when the sampler next attempts to draw from it, guaranteeing that the next draw of the EoR field, $\mathbf{e}$, has a very low variance. In turn, this makes it much more likely for a small value of the band power $P(\tau)$ to be drawn in the following iteration, and so on. The Gibbs sampler then gets `stuck', potentially for many iterations, in a region of the parameter space where $\mathbf{e}$ and $P(\tau)$ are both small. Approaches such as deterministically rescaling the random draws can help avoid this issue \citep{2016ApJ...820...31R}, but we do not pursue this further, and rely on the truncated prior bounds instead.

While a truncated inverse Gamma prior is adopted here to minimise assumptions about the underlying 21-cm signal and to isolate the impact of the systematics model, the framework is not restricted to this choice. More physically motivated priors, for example those informed by numerical simulations or emulator predictions of the 21-cm power spectrum, could be incorporated within the same Bayesian framework. Such priors may also help to improve sampling efficiency.

\section{Results}\label{sec:results}

In this section, we present results obtained for different simulated test cases, following the approach in Sect.~\ref{sec:sysmodel}. The three cases are formulated based on systematic artefacts found in real data \citep{Kern2020b} and simulations \citep{Aguirre2022}, and were shown in Fig.~\ref{fig:test_data_dlfr}. The purpose of this exercise is to demonstrate the response of the sampler to artefacts placed at various locations in the delay-fringe rate space. The systematics model can be made more complex by adding more modes, and the model freedom restricted by choosing different prior hyperparameters, but we restrict ourselves to only the three cases in what follows.

In Case~\RNum{1}, we introduce systematic effects to four delay-fringe rate mode pairs situated around the same fringe-rate as the sky ($f \approx 0$~mHz), and at low delay modes ($\tau \approx 300-600$~ns) that are just outside the foreground-dominated region. We anticipate that these would be the most challenging to model due to substantial overlap with the foregrounds.

In Case~\RNum{2}, the effects are introduced to four mode pairs situated at the same fringe rates, but substantially higher delay modes ($\tau \approx 1000-1300$~ns), away from the foreground-dominated delays. This case corresponds to simplified versions of the cable reflections seen in \cite{Kern2020b}.

In Case~\RNum{3}, the systematics are introduced to four mode pairs again, situated at the same low delay modes as in Case~\RNum{1}, but now at fringe-rates of $f \approx 6$~mHz, significantly separated from the sky components. This case is more akin to a cross-coupling systematic, but also serves to test how much separation of the systematics from the sky signal in fringe rate affects the recovery.

For each case, we draw $250,000$ samples with the Gibbs sampler, rejecting the initial $10\%$ of samples to account for burn-in effects. We include an uncorrelated Gaussian noise model with known variance $\mbf{N}$, set to have an RMS value of $0.344$ times the RMS value of the EoR signal, yielding an SNR of about 3.

\subsection{Sky component recovery}\label{subsec:res_vis}

Fig.~\ref{fig:result_waterfalls} shows a comparison of the true (simulated) sky visibilities and the recovered (posterior predictive distribution) mean, plotted in delay-fringe rate space. The standard deviation of the posterior predictive distribution, and the residuals (i.e. the difference between the true model and posterior predictive mean), are also shown. In all three cases, the true sky is recovered reasonably well. The imprint of the systematic effects can be seen clearly in the residual maps and the standard deviation maps however, particularly in Cases~\RNum{1} and \RNum{3}. This can largely be attributed to correlations between the foreground and systematics models, which have sufficiently overlapping structures in delay space that they cannot be easily or uniquely separated. This also slows down convergence of the sampler, and it is plausible that returning many more samples would allow better exploration of the correlated/degenerate directions in the parameter space. This is further demonstrated by the lower amplitude of the standard deviation and residuals in the systematics-affected regions in Case~\RNum{2}, which are well-separated from the foregrounds. We discuss this further in Sect.~\ref{subsec:corr} and establish that in Case~\RNum{2}, systematics and foregrounds are indeed less correlated in the posterior distribution. 

Fig.~\ref{fig:comp_errors} (upper panel) shows the DPS of the recovered sky components individually for Case~\RNum{3}. The difference between the recovered (mean) EoR DPS and true EoR DPS is shown in the lower panel. We observe that the EoR delay spectrum is recovered to approximately the noise level. The only outlying residual fluctuations are present at lower delay modes bordering the foreground dominated region, which is expected due to the difficulty of separating low-$\tau$ EoR modes from smooth foregrounds.

Returning to the correlations between the different components of the model, Fig.~\ref{fig:res_comp} shows residuals between the true and posterior mean models for each component as a function of frequency and time (LST), compared with the simulated noise. On large spectral and temporal scales, the EoR and foreground component residuals show a clear anti-correlation, which signifies that one is absorbing power from the other. Without tighter priors or direct modelling of the temporal (LST) structure of each component, there is nothing to distinguish them from one another on these scales, and so some degree of (anti-)correlation is inevitable.

\subsection{Systematic component recovery}\label{subsec:res_sys}

\begin{figure*}
    \centering
    \includegraphics[width=\linewidth]{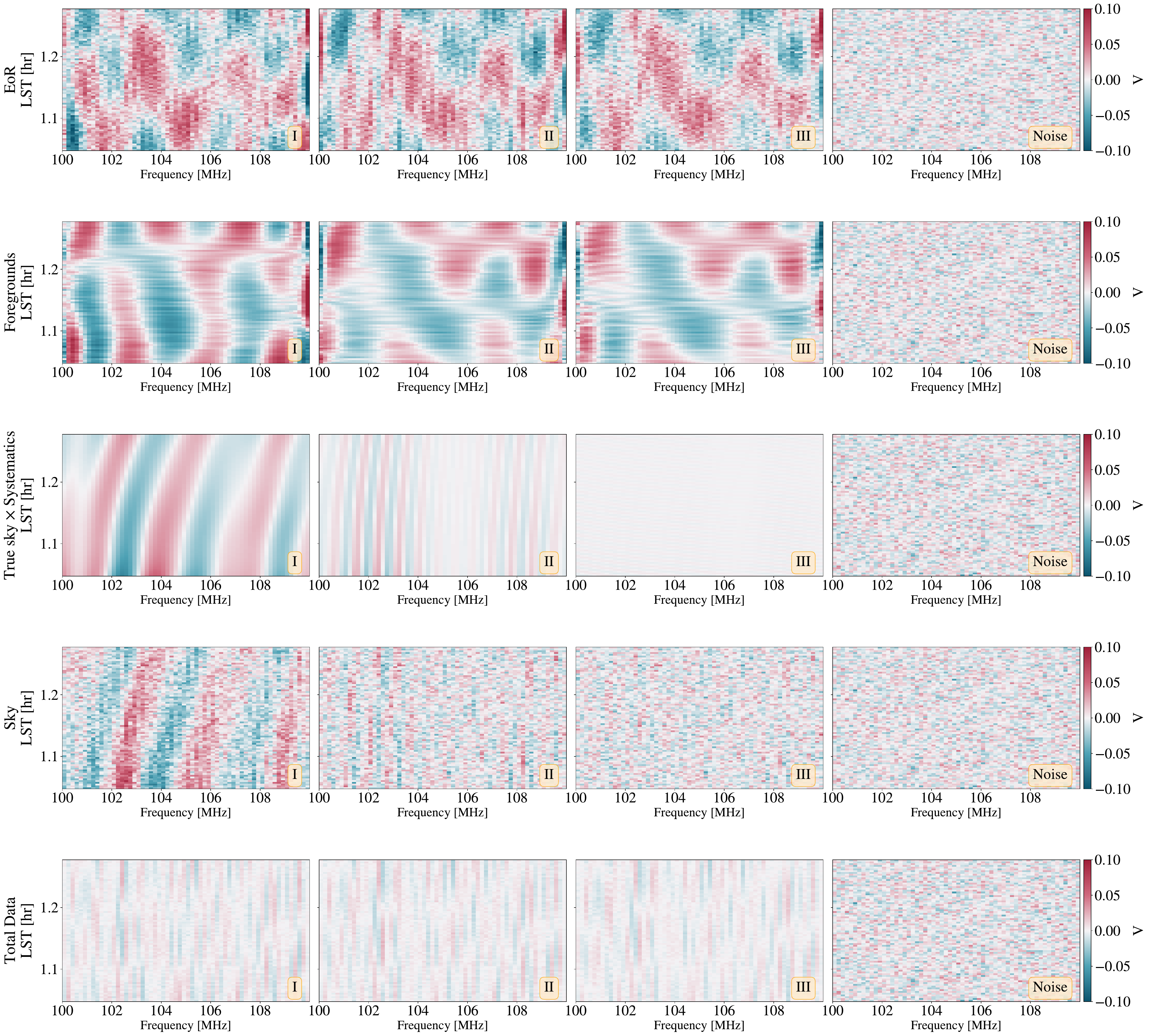}
    \caption{The residuals of the component visibilities, calculated as the difference between the true model and the posterior mean model. The residuals are shown as a function of frequency and time. Case \RNum{1} shows slightly higher residuals owing to the strong correlations between the systematics artefacts and the foregrounds. There are some structures present in the EoR and foreground residuals, which is also expected since we are not modelling the time correlations in our data. The noise panels are identical for each row, and for each case.}
    \label{fig:res_comp}
\end{figure*}

We also study the recovery of the systematics component in Fig.~\ref{fig:res_comp}, which shows an anti-correlation in the residuals between the total sky model and the systematics model (multiplied by the true sky model to make the comparison more intuitive). This is most apparent for Case~\RNum{1}, which has a clear residual systematic that is a few times larger than the noise level. Case~\RNum{2} has a much smaller residual systematic that is around the noise level, while the residual is not clearly visible for Case~\RNum{3} on the same colour scale.

\begin{figure*}
    \centering
    \includegraphics[width=0.8\linewidth]{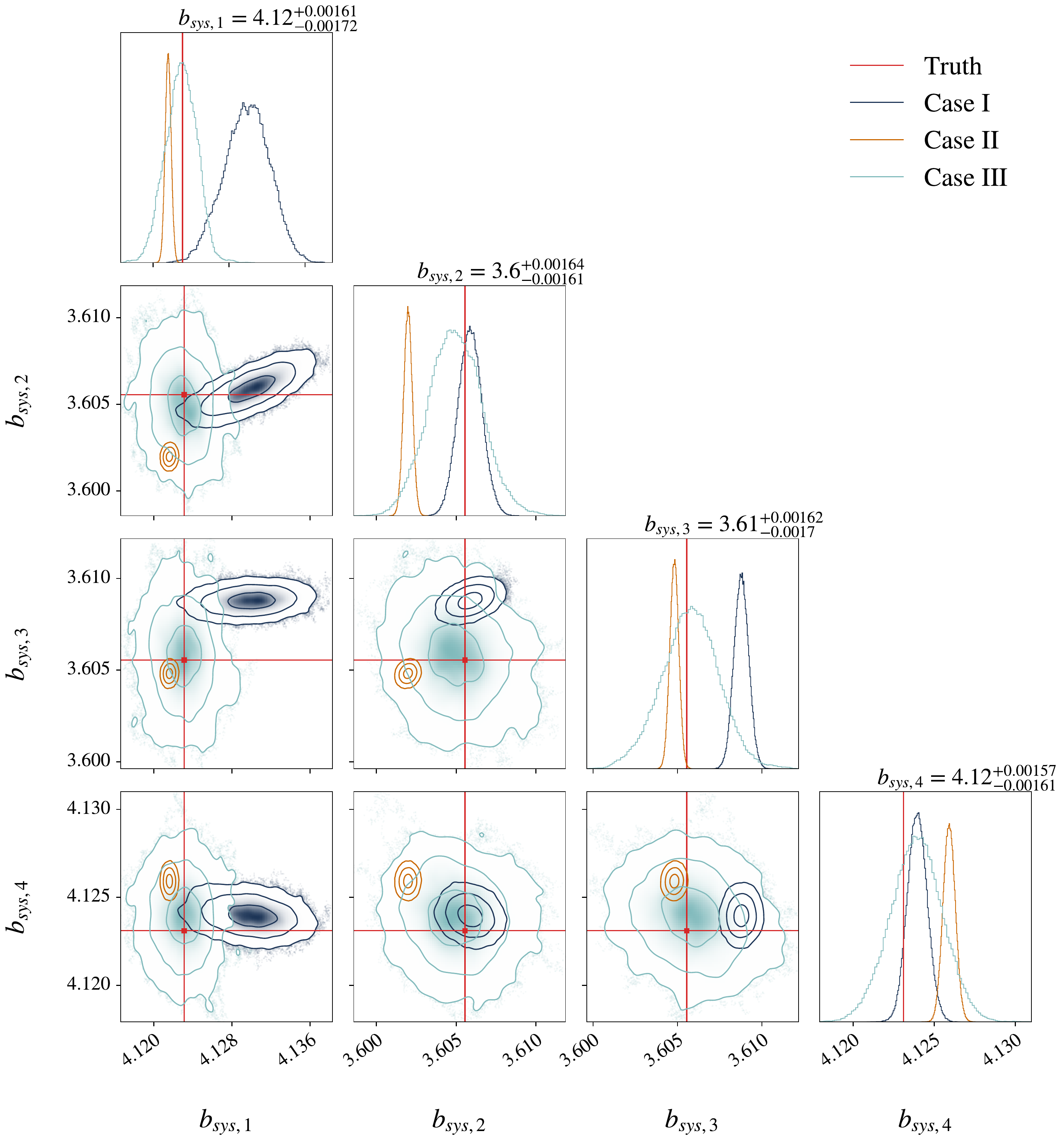}
    \caption{Corner plot of the absolute value of the systematics parameters, $\mathbf{b}_\textup{sys}$, for all three cases. The outer contours represent a $3\sigma$ bound around the mean value. The red crosshairs indicate the truth value of each parameter (which is actually composed from separate real and imaginary parameters).}
    \label{fig:corner_plots}
\end{figure*}

Further insight into these results can be gained from Fig.~\ref{fig:corner_plots}, which shows a corner plot of the absolute values of the recovered systematics parameters. Because the parameter values were all chosen to be the same across all three cases, the truth value (shown by the red crosshair) is the same for all cases in each panel. These plots show excellent recovery of all four (complex) parameters for Case~\RNum{3}, with the true values within one estimated standard deviation of the recovered values for all of them. For Case~\RNum{2}, the recovery is more confident -- the contours are significantly narrower -- but they are offset from the true value by several standard deviations in most cases. Case~\RNum{1} is more mixed, with broader contours, but a mix of correct and incorrect recoveries. We will investigate the reasons for, and implications of, the incorrect recoveries in the following sub-sections.

It is worth noting here that the multiplicative systematics term is modelled as $\mbf{y}=1+\mbf{H}\mbf{b_\tup{sys}}$, where $\mbf{H}$ is a linear operator containing a {\it reduced} set of delay-fringe rate (Fourier) modes. This requires the Fourier modes affected by the systematics to be identified a priori, e.g. by inspecting the delay-fringe rate waterfalls or by performing a first-pass analysis using other tools.
In our test cases, since we are artificially injecting systematics at specific Fourier modes, we already have the accurate mode information required to form $\mbf{H}$. This enables us to test the sampler in a fairly optimistic scenario, where the affected modes are already known. In reality, a broader set of modes around identified systematics features would probably need to be included.

\subsection{Recovery of the EoR delay power spectrum}\label{subsec:res1}

The above results show that the total sky model (foreground plus EoR) and systematics models are recovered reasonably well, but not perfectly, with some discrepancies that are statistically significant. The problem is amplified by the high signal-to-noise ratio we have assumed in our tests; current data have a much lower SNR, making it easier for such discrepancies to be buried in the noise. Nevertheless, it is important to establish what effect these discrepancies have on the recovered EoR delay power spectrum, which is the main scientific target of the relevant 21\:cm cosmology surveys. 

\begin{figure*}
    \centering
    \includegraphics[width=\linewidth]{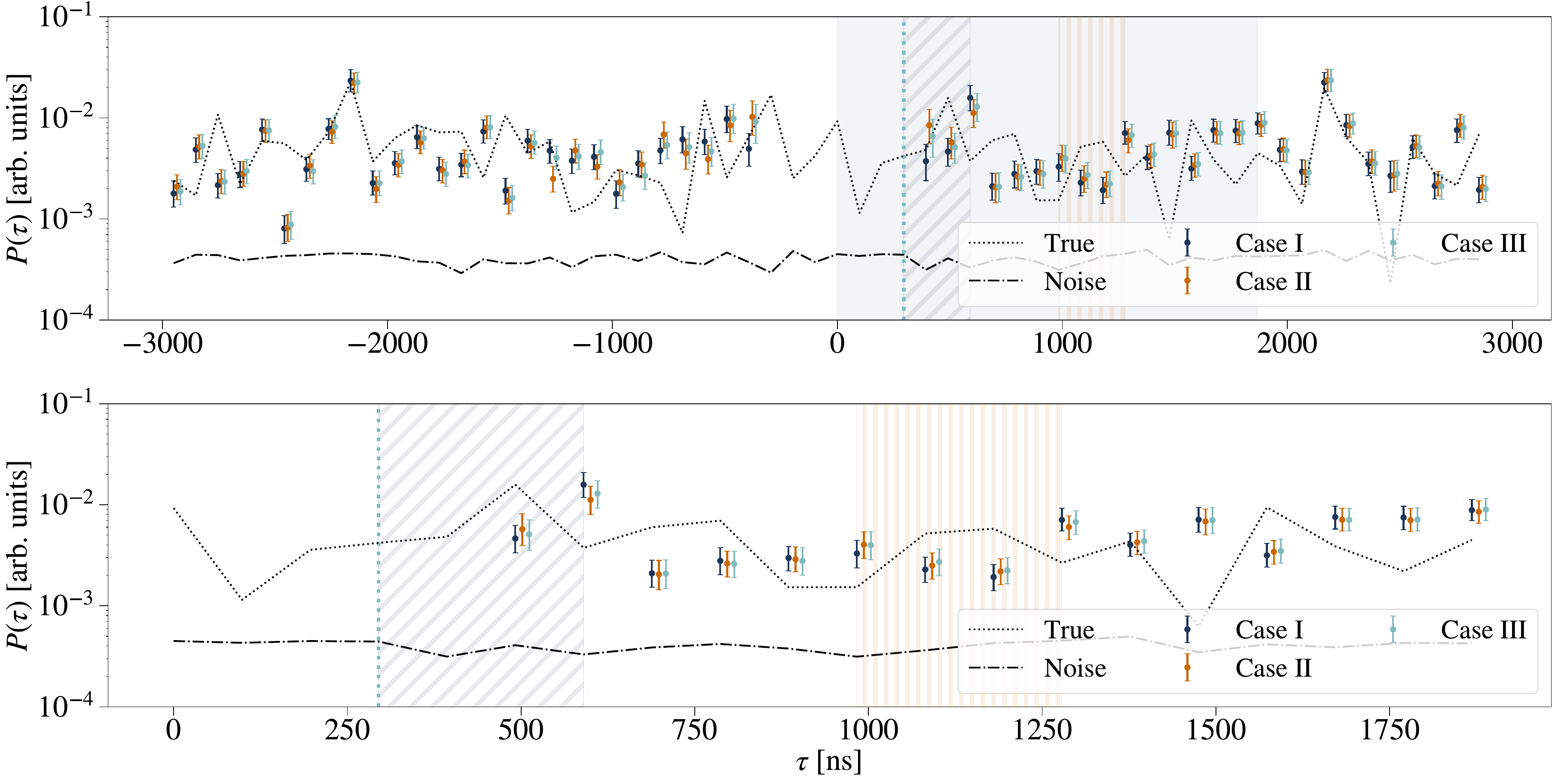}
    \caption{This plot shows the recovery of the delay power spectrum (DPS) from the solver for the three test cases. The top panel shows the recovered DPS when compared against the true EoR DPS and the bottom panel shows a zoomed in version of the same plot for better visualisation. The solid shaded region in the top panel indicates the scope of the zoomed-in plot. The black dotted line in the indicates the true EoR DPS and the coloured dots indicate the mean of the sampled DPS for each case. The error bars correspond to $95\%$ credible intervals. The diagonally shaded sky blue region (at delay modes $< 1000$ ns) indicates the delay modes where the systematics were injected in Case~\RNum{1}, the vertically shaded light orange region (at delay modes $> 1000$ ns) similarly refers to Case~\RNum{2}, and the dotted vertical line refers to Case~\RNum{3} where, we inject the systematics at the same delays as Case~\RNum{1}, but at fringe rate $\approx 6 \, \mathrm{mHz}$.}
    \label{fig:dps}
\end{figure*}

Fig.~\ref{fig:dps} shows recovery of the EoR DPS for each of the three test cases compared against the true EoR DPS. 
We have omitted delays in the centre of the delay range, as this corresponds to the foreground dominated region where strong (posterior) correlations between the EoR and foreground signals are expected to hamper recovery regardless. We have suppressed this region, with suitable priors chosen to avoid the EoR component absorbing an unrealistic amount of the foreground power, and do not attempt EoR signal recovery here. This issue was discussed in more detail by \cite{Burba2024}.

The upper panel of the figure shows that the EoR DPS is recovered quite well across the remaining delay range, and that the recovery is quite similar for all three cases. For most delay bins, the three cases produce very similar results, to well within a single standard deviation. A few delay bins show larger uncertainty for some of the cases, and more scatter between the cases. This is most notable at around $\pm 500$~ns, where the Case~\RNum{1} values are scattered a little higher than the others, and the Case~\RNum{2} uncertainties are substantially larger. 

Fig.~\ref{fig:dps} also shows a zoomed-in region of delay modes in the bottom panel. Here, it is clear that the recovery of the DPS at higher delay modes outside of the foreground dominated region is largely independent of the location of the systematic artefacts, denoted by shaded/hatched regions. Again, there are discrepancies, with some DPS values being multiple standard deviations away from the true value. We anticipate that this is due to the same underestimation of uncertainties that was noted by \cite{Burba2024}, which is caused by not accounting for the temporal correlations of the EoR signal.

\subsection{Correlations and their impact on sampling efficiency}\label{subsec:corr}

Despite their ability to handle very high-dimensional parameter spaces, Gibbs samplers are still Markov Chain methods that randomly explore the parameter space in a chain of steps, each one depending on the position of the last. This leads to correlations between samples drawn at each step. The effective number of independent samples drawn from the posterior is therefore less than the actual number of samples drawn, depending on the strength of this correlation effect.

The efficiency of the Gibbs sampler depends on correlations between parameters drawn from different conditional distributions within the Gibbs scheme. For parameters sampled from the same conditional distribution, their correlation/degeneracy direction is explored directly and efficiently if it is an exact sampling step, like GCR. This type of parameter correlation should not affect the correlation length of the chain appreciably. For parameters with strong correlations that are drawn from different conditional distributions however, the exploration of the correlated/degenerate direction is slow as the chain must iterate between steps in one parameter and then the other. This effect is described in some detail in \cite{2016ApJ...820...31R}. The downside is that strong correlations between foreground parameters (in one Gibbs step) and systematics parameters (in another) can lead to long correlation lengths, making the exploration of the posterior distribution inefficient.

The correlation of the chain can be evaluated by studying the correlation between the samples, which can be quantified by the integrated correlation time and the effective sample size (ESS). The integrated correlation time gives us an estimate of how many iterations we need to execute to obtain consecutive {\it independent} samples, while the ESS gives us an estimate of how many independent samples we have obtained in effect from our total number of iterations \citep{bayesian1995}.

\begin{table}
\centering
\resizebox{\columnwidth}{!}{
\begin{tabular}{c|c|cccc}
\hline
Metric & Case 
& $b_{\mathrm{sys},1}$ 
& $b_{\mathrm{sys},2}$ 
& $b_{\mathrm{sys},3}$ 
& $b_{\mathrm{sys},4}$ \\
\hline

\multirow{3}{*}{Integrated Correlation Time }
& Case~\RNum{1} 
& 1375.5 & 1046.4 & 99.1 & 126.2 \\
& Case~\RNum{2} & 25.9 & 20.5 & 19.6 & 38.1 \\
& Case~\RNum{3} & 1278.4 & 987.9 & 433.5 & 114.5 \\

\hline

\multirow{3}{*}{ESS}
& Case~\RNum{1} & 182 & 239 & 2522 & 1981 \\
& Case~\RNum{2} & 9644 & 12189 & 12724 & 6555 \\
& Case~\RNum{3} & 195 & 253 & 577 & 2183 \\

\hline

\multirow{3}{*}{Execution time (h)}
& Case~\RNum{1} & \multicolumn{4}{c}{7.43} \\
& Case~\RNum{2} & \multicolumn{4}{c}{7.42} \\
& Case~\RNum{3} & \multicolumn{4}{c}{7.36}\\

\hline
\end{tabular}
}
\caption{Integrated correlation times, effective sample sizes (ESS), and execution times for all parameters and cases.}
\label{tab:corr_values}
\end{table}

The ESS and integrated correlation time are given in Table~\ref{tab:corr_values} for the three cases we studied. While the posterior for the systematics parameters is explored reasonably efficiently for Case~\RNum{2}, excessive correlation times are found for Cases~\RNum{1} and \RNum{3}. Cases~\RNum{1}, \RNum{2} and \RNum{3} each returned the full $250,000$ samples from the Gibbs sampler over a reasonable wall-clock execution time of about 7.4~hours, yet a little under 200 independent samples of the first systematics parameter are returned for Case~\RNum{1}. This corresponds to a very low efficiency, and is likely to have affected the convergence of the chains, which may have contributed to the discrepancies seen in the recovered values of some of the parameters. These differences make Case~\RNum{2} stand out from the other two cases in both the corner plot in Fig. \ref{fig:corner_plots} and in the DPS comparison in Fig. \ref{fig:dps}. 

\begin{figure*}
    \includegraphics[width=\textwidth]{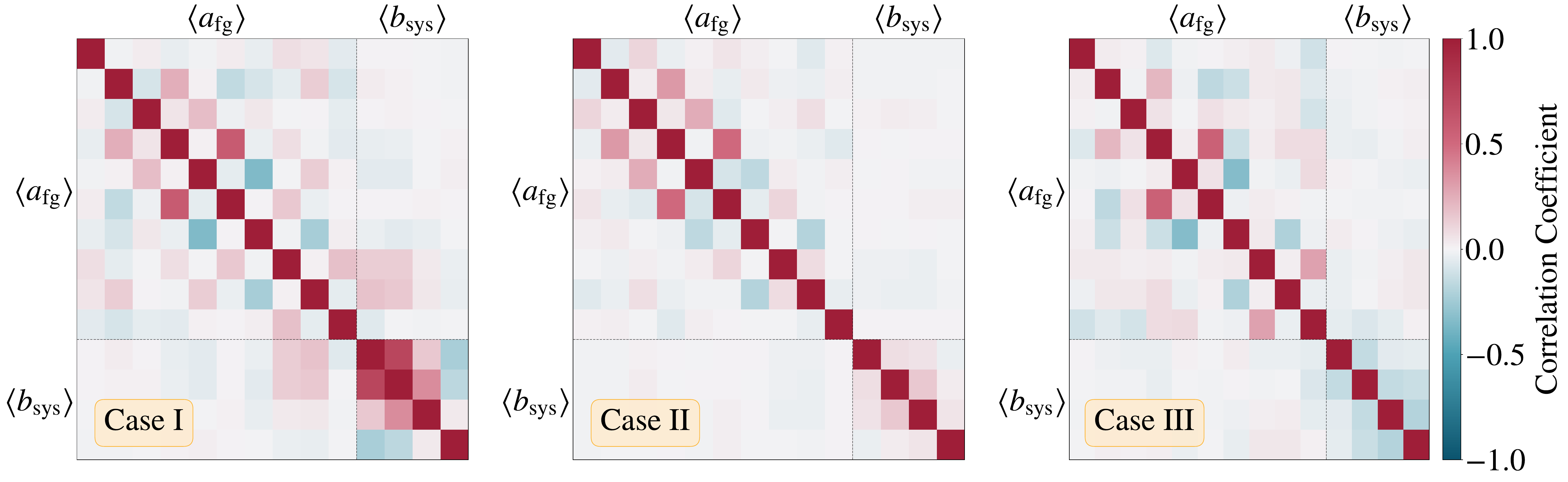}
    \caption{Pearson's correlation coefficient $\mbf{r}$ between the foreground parameters $\mbf{a}_\tup{fg}$ and the systematics parameters $\mbf{b}_\tup{sys}$ for one time sample have been shown here for all the three test cases. The non-zero correlation coefficients between these two sets of parameters in the off-diagonal positions generate artefacts of the systematics contaminations when the sky visibilities are sampled. They also lengthen the correlation lengths of our sampler, making each sample less efficient, particularly in cases where the same delay modes are shared by both components, as in Cases \RNum{1} and \RNum{3}.}
    \label{fig:corr}
\end{figure*}

It is possible to try to identify which parameters give rise to the correlations that are likely to be causing this behaviour. Fig.~\ref{fig:corr} shows the posterior marginal covariance of the foreground amplitudes $\mbf{a}_\tup{fg}$ and systematics parameters $\mbf{b}_\tup{sys}$. We have visualised these correlations for the values of $\mbf{a}_\tup{fg}$ for one time sample for each of our test cases. There are non-trivial off-diagonal correlations between the two sets of parameters in Case~\RNum{1} and \RNum{3} that are essentially absent for Case~\RNum{2}. 

This is perhaps unsurprising for Case~\RNum{1}, as there is a direct overlap in delay-fringe rate space between the foregrounds and systematics. In Case~\RNum{3} they are disjoint however. In this latter case, we anticipate that the strong correlations arise because the foreground model is fitted independently at each time (LST); there is no attempt to model its temporal structure. This means that the foregrounds at each time sample can independently absorb systematics structure in delay space in a way that conspires to induce foreground (and EoR) residuals with a similar fringe rate structure to the systematics. Fig.~\ref{fig:res_comp} gives support to this hypothesis, as the residuals for the total sky component and systematics are seen to be anti-correlated on large spectral and temporal scales. The systematics residuals are negligible for Case~\RNum{3} in Fig.~\ref{fig:res_comp}, so any (anti-)correlation between them and the total sky component is not seen in this figure.  

In Case~\RNum{2}, there are no such correlations. However, despite the short autocorrelation times and large effective sample size indicating that the posterior has been well explored, the recovered systematics parameters remain biased for this case. We attribute this to a different mechanism. Since the foreground amplitudes $\mathbf{a}_{\text{fg}}$ are fitted independently at each LST, they are able to conspire across time samples to jointly absorb some of the systematics structure present in the overlapping fringe rates, even though the systematics themselves are well-separated from the foregrounds in delay. This differs from the direct delay-space overlap responsible for the long correlation times seen in Cases~\RNum{1} and \RNum{3}, and highlights that foreground-systematics correlations can arise through more than one channel in this framework. We also needed to increase the number of Gibbs samples from an initial $10,000$ to $250,000$ to improve the convergence of the sampler.

Our findings make an optimistic case for using this method to model cable reflections, which occur at higher delay modes (similar to Case~\RNum{2}), but the long correlation time issue is problematic in cases where systematics appear at foreground-dominated delay modes, even if they are well-separated in fringe rate.

There are several possibilities for ameliorating this behaviour. One is to simply impose stronger priors; if a first-pass analysis is able to identify the approximate systematics parameter values with reasonable accuracy, smaller prior covariance values can be assigned to $\mbf{B}$, which should break the strong correlations/degeneracies between the parameters sufficiently well to allow more rapid exploration of the parameter space without being overly restrictive. Another option is to implement a proper model for the temporal correlations of the foregrounds, which are currently treated as independent between time samples. This was already remarked upon as an important future development for the EoR DPS modelling by \cite{Burba2024}, but would require some re-engineering of the \code{hydra-pspec} code. Finally, the joint rescaling approach outlined in \cite{2016ApJ...820...31R} \citep[see also][]{2009ApJ...697..258J} offers the possibility of removing the bottleneck if a suitable rescaling can be identified.

\subsection{Combined systematics recovery}\label{subsec:res_combined}

Having examined the sampler's response to each systematics configuration in isolation, we now consider a combined test case in which the artefacts from Cases~\RNum{1}, \RNum{2}, and \RNum{3} are injected simultaneously. In this test case, $\mbf{b}_\mathrm{sys,1} - \mbf{b}_\mathrm{sys,4}$ are located at the Fourier modes used in Case~\RNum{1}, $\mbf{b}_\mathrm{sys,5} - \mbf{b}_\mathrm{sys,8}$ are located at the Fourier modes of Case~\RNum{2}, and $\mbf{b}_\mathrm{sys,9} - \mbf{b}_\mathrm{sys,12}$ are located at the Fourier modes used in Case~\RNum{3}.

This also gives us an opportunity to further study the correlations between foregrounds and systematics discussed in Sect.~\ref{subsec:corr}. We again draw $250,000$ samples with the Gibbs sampler, rejecting the initial $10 \%$ as burn-in, using the same noise model as in Sect.~\ref{sec:results} (RMS $0.344\times$ the RMS of the EoR signal, SNR $\approx 3$).

This is a more realistic test of the sampler's performance, since real data are unlikely to contain a single, isolated systematic of the kind explored in Sects.~\ref{subsec:res_vis}--\ref{subsec:corr}, and since Sect.~\ref{subsec:corr} established that Cases~\RNum{1} and \RNum{3} in particular induce non-trivial correlations between the foreground and systematics parameters. The purpose of this combined case is to establish whether these effects interact, and whether the degradation in recovery accuracy and sampling efficiency seen when systematics overlap the foregrounds in delay (Case~\RNum{1}) or when they share fringe-rate structure with the foregrounds despite delay separation (Case~\RNum{3}) simply adds when both mechanisms are present at once, or whether a more complex, non-additive interaction emerges.

\begin{figure*}
    \centering
    \includegraphics[width=\linewidth]{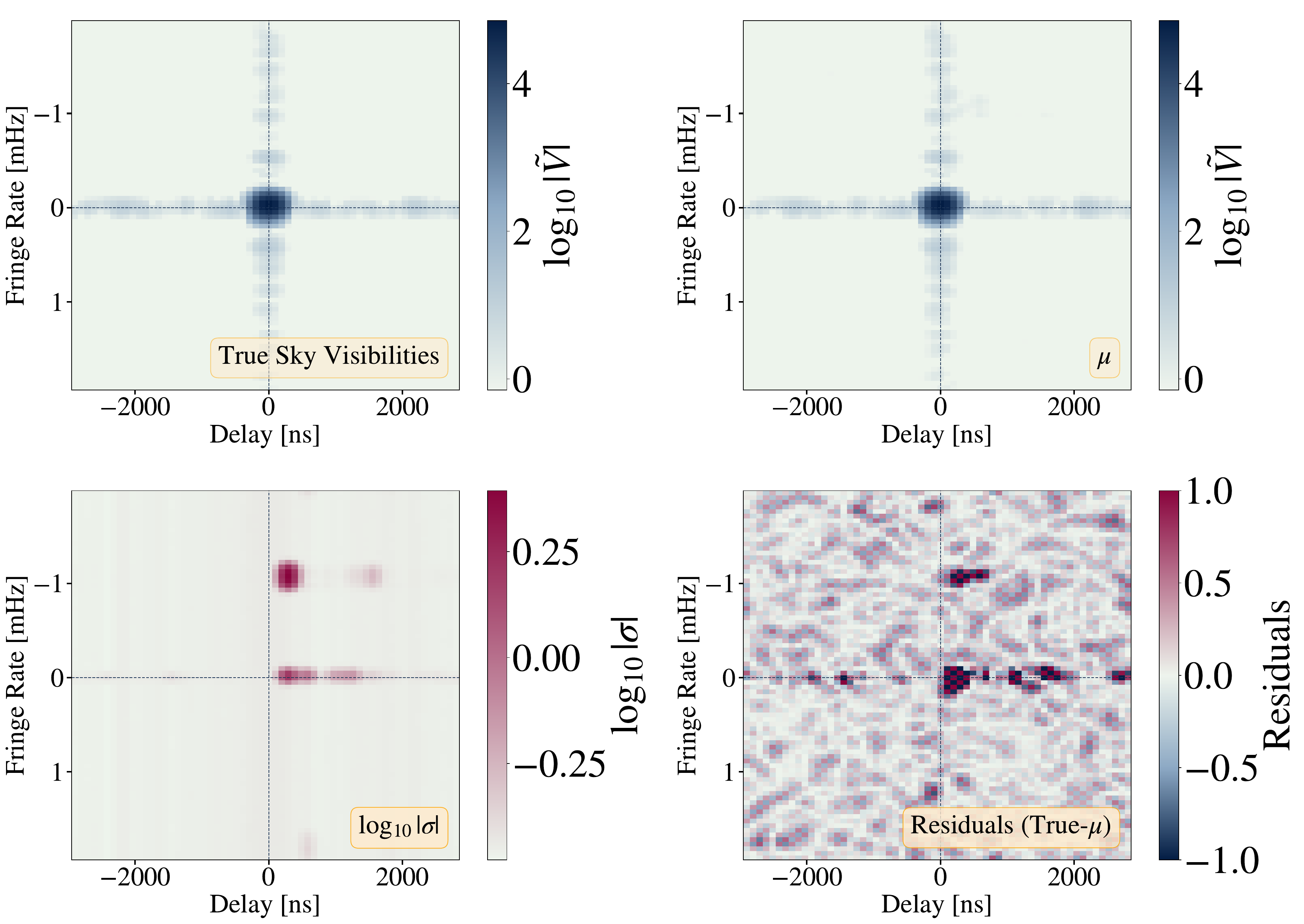}
    \caption{Recovery of the sky visibilities for the combined systematics case, shown in delay-fringe rate space. This is the analogue of Fig.~\ref{fig:result_waterfalls} for the test in which all twelve systematics modes are injected simultaneously. \textit{Top left}: The true sky visibilities, containing the EoR and foreground components. \textit{Top right}: the posterior predictive mean $\mu$, formed by summing the EoR and foreground samples at each Gibbs iteration and averaging over iterations. \textit{Bottom left}: the standard deviation $\sigma$ of the same samples, in which the imprint of the injected systematics is visible as localised increases at the affected delay-fringe rate modes. \textit{Bottom right}: the residuals between the true sky and the posterior predictive mean, normalised by $\sigma$. Note that, unlike Fig.~\ref{fig:result_waterfalls}, the standard deviation is shown on a linear scale and the residuals are normalised, so the two figures should be compared qualitatively rather than panel by panel.}
    \label{fig:caseiv_sky}
\end{figure*}

\begin{figure}
    \centering
    \includegraphics[width=\columnwidth,keepaspectratio]{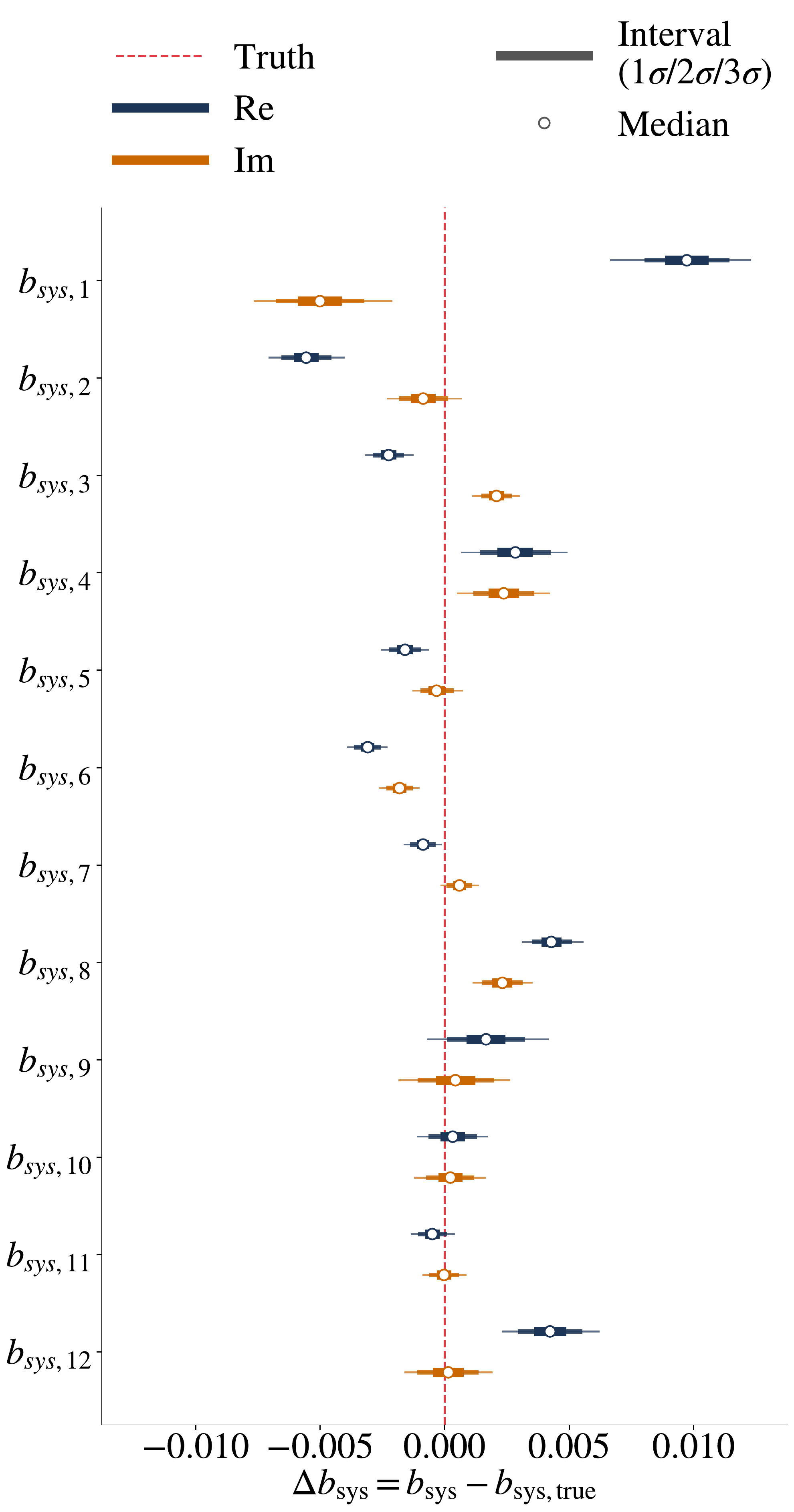}
    \caption{Offset of the recovered systematics parameters from their true values, $\Delta b_\mathrm{sys} = b_\mathrm{sys} - b_\mathrm{sys}^\mathrm{true}$, for the combined systematics case, shown separately for the real (dark blue) and imaginary (orange) components of each of the twelve parameters. For each component, the thick, medium, and thin horizontal bars show the $1\sigma$, $2\sigma$, and $3\sigma$ credible intervals respectively, with the open circle marking the posterior median. The dashed red line marks zero offset (perfect recovery). Parameters whose credible intervals do not overlap this line are biased away from their true value at the corresponding confidence level — consistent with the reduced recovery accuracy noted for the combined case in Sect.~\ref{subsec:res_combined}.}
    \label{fig:multi_corner_plots}
\end{figure}

The true sky is recovered with a larger residual than in any of the individual test cases. The RMS residual between the true sky and its posterior predictive mean is $0.55$ times the RMS of the true EoR signal, against $0.41$, $0.30$ and $0.27$ for Cases~\RNum{1}, \RNum{2} and \RNum{3} respectively, so that only Case~\RNum{1} and the combined case exceed the noise level of $0.344$. The combined residual is $0.96$ times the quadrature sum of the three individual residuals, however, so the contaminations add rather than compound. This is shown in Fig.~\ref{fig:caseiv_sky}, the combined-case analogue of Fig.~\ref{fig:result_waterfalls}. The residual and standard deviation characteristics closely resemble those of the previous cases, with imprints appearing at the delay-fringe rate locations corresponding to where the systematics were injected. This is consistent with the expectation from Sect.~\ref{subsec:corr}. The ordering of the individual residuals identifies delay overlap with the foregrounds as the more damaging of the two mechanisms: Case~\RNum{1}, whose modes sit at foreground-dominated delays, leaves the largest residual of the three, and Case~\RNum{3}, separated from them in delay, the smallest. The systematics parameters themselves, shown in Fig.~\ref{fig:multi_corner_plots}, are recovered with qualitative patterns similar to the previously discussed cases, with the foreground-adjacent delay positions once again suffering the most in terms of accuracy.

The standard deviation map in Fig.~\ref{fig:caseiv_sky} makes the multiplicity of the contamination explicit. Localised increases in the posterior standard deviation appear at each group of injected modes: at the Case~\RNum{1} delays, $\tau \approx 300-600$~ns, at zero fringe rate; at the Case~\RNum{2} delays, $\tau \approx 1000-1300$~ns, also at zero fringe rate but visibly weaker; and at the Case~\RNum{3} modes, which share the Case~\RNum{1} delays but are offset in fringe rate. The ordering of these features by amplitude mirrors the ordering of the correlation times in Table~\ref{tab:multi_corr_tab}. The modes that overlap the foregrounds in delay carry the largest posterior uncertainty, while the well-separated Case~\RNum{2} modes are the most tightly constrained, which is the same pattern seen for the individual cases in Fig.~\ref{fig:result_waterfalls}. The normalised residual map in the lower right panel is largely noise-like, but retains coherent positive features at these same locations. Some systematics power therefore remains unabsorbed even once the enlarged posterior uncertainties are taken into account, consistent with the degraded parameter recovery reported above and with the DPS behaviour discussed below.

\begin{figure*}
    \centering
    \includegraphics[width=\linewidth]{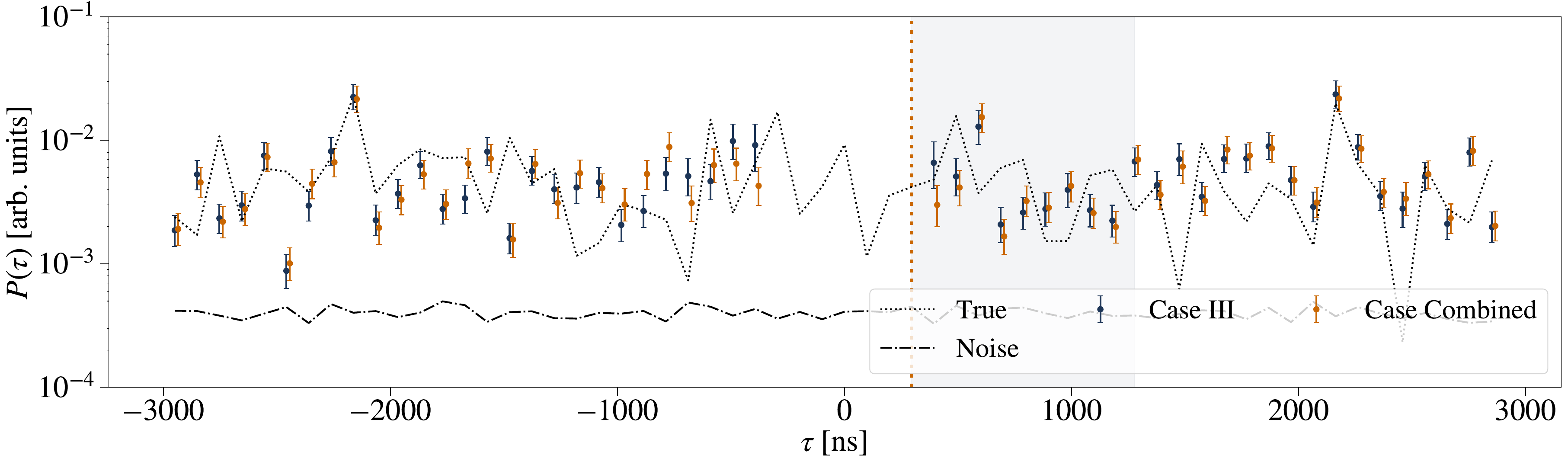}
    \caption{Recovery of the delay power spectrum (DPS) for the combined systematics case, compared against the true EoR DPS, compared against the recovery from Case~\RNum{3}. The dotted line indicates the delay mode where systematics are placed in Case~\RNum{3}. The shaded region indicates the delay ranges over which systematics have been placed in the combined case. This essentially spans all the delays covered in the three test cases.}
    \label{fig:multi_dps}
\end{figure*}

Fig.~\ref{fig:multi_dps} shows that the recovered EoR DPS for the combined case is essentially indistinguishable from that of Case~\RNum{3}, which we chose as our ``control'' case from the three individual cases. Across the 53 delay bins outside the foreground-dominated region, the median width of the $95\%$ credible interval differs by less than $1\%$ between the two, and the median deviation of the recovered DPS from the true DPS, normalised by the posterior standard deviation, changes only from $4.35\sigma$ to $4.43\sigma$. The number of bins discrepant at more than $2\sigma$ is 40 of 53 for Case~\RNum{3} and 39 of 53 for the combined case. Injecting systematics at twelve Fourier mode pairs rather than four therefore neither widens the posterior on the EoR DPS nor moves it further from the truth, at the level these tests can resolve. This should be set against the systematics parameters themselves (Fig.~\ref{fig:multi_corner_plots}), which \emph{are} recovered less accurately in the combined case: the degradation is confined to $\mbf{b}_\mathrm{sys}$ and does not propagate into the recovered EoR power spectrum.

We stress that this is a statement about the combined case \emph{relative} to the individual ones. Neither recovers the EoR DPS to within its own quoted uncertainties: the median normalised deviation is $\approx 4\sigma$ in both, with 33 and 36 of the 53 delay bins respectively lying more than $3\sigma$ from the true value, corresponding to median fractional deviations of $0.63$ and $0.66$. This is the same underestimation of the uncertainties noted in Sect.~\ref{subsec:res1}, which we attribute to the omission of the temporal correlations of the EoR signal from our model, following \cite{Burba2024}. It affects the individual cases and the combined case alike, and is therefore not a consequence of combining systematics configurations.

Table~\ref{tab:multi_corr_tab} shows integrated correlation times and effective sample sizes across the twelve systematics parameters. Notably, the worst-case integrated correlation time ($\tau$) and worst-case ESS for the combined case are both substantially improved relative to the worst-case values seen across the individual cases (cf.\ Table~\ref{tab:corr_values}). These improvements in sampling efficiency do not come at the cost of the DPS recovery, which is statistically indistinguishable from that of Case~\RNum{3} by the measures quantified above.

The worst-case $\tau$ is 330.5 for $\mbf{b}_\mathrm{sys,1}$, the same parameter that performed worst in Case~\RNum{1}. Rather than degrading, $\tau$ decreases from 1375.5 to 330.5, a reduction by a factor of over four, while the worst-case ESS increases from 182 to 756. This indicates that combining the three configurations does not compound the correlation-driven sampling inefficiency identified in Sect.~\ref{subsec:corr}. The runtime for $250{,}000$ samples increases from $\sim7.4$ hours for the individual cases to 9.29 hours ($0.13$~s per iteration), a factor of $\sim1.25$. This modest increase is expected because the dominant GCR sampling cost is unchanged, while the systematics block grows from 4 to 12 parameters and remains comparatively cheap to sample.

\begin{table}
\centering
\resizebox{\columnwidth}{!}{
\begin{tabular}{c c SS}
\toprule
Source Case & Parameter (complex) &
\multicolumn{1}{c}{Integrated Correlation Time} &
\multicolumn{1}{c}{ESS} \\
\midrule
Case~\RNum{1} & $b_{\mathrm{sys},1}$ & 330.5 & 756 \\
& $b_{\mathrm{sys},2}$ & 88.3 & 2831 \\
& $b_{\mathrm{sys},3}$ & 53.9 & 4637 \\
& $b_{\mathrm{sys},4}$ & 91.2 & 2742 \\
\hline
Case~\RNum{2} & $b_{\mathrm{sys},5}$ & 43.1 & 5803 \\
& $b_{\mathrm{sys},6}$ & 32.7 & 7652 \\
& $b_{\mathrm{sys},7}$ & 33.3 & 7511 \\
& $b_{\mathrm{sys},8}$ & 48.2 & 5185 \\
\hline
Case~\RNum{3} & $b_{\mathrm{sys},9}$ & 256.9 & 973 \\
& $b_{\mathrm{sys},10}$ & 270.3 & 925 \\
& $b_{\mathrm{sys},11}$ & 227.6 & 1099 \\
& $b_{\mathrm{sys},12}$ & 286.6 & 872 \\
\hline
\multicolumn{4}{c}{Execution time: 9.29 hours} \\
\bottomrule
\end{tabular}
}
\caption{Integrated correlation times, effective sample sizes (ESS), and
execution time for the combined systematics case.}
\label{tab:multi_corr_tab}
\end{table}

These results paint a mixed picture. While the correlation-driven sampling inefficiency identified in Sect.~\ref{subsec:corr} does not compound when systematics configurations are combined — indeed, the worst-case correlation time improves by more than a factor of four — the accuracy with which the systematics parameters themselves are recovered nonetheless degrades relative to the individual cases, even as the recovered EoR DPS does not. This suggests that sampling efficiency and parameter recovery accuracy are not simply two faces of the same underlying difficulty, and that a well-mixing chain does not guarantee an accurate one when multiple systematics configurations are present simultaneously. It also suggests that the EoR delay power spectrum, the quantity of scientific interest, is the more robust of the two to this form of contamination.

\section{Discussion and Conclusions}
\label{sec:conclusions}

Separating the redshifted 21cm signal from bright radio foregrounds is a challenging task. Instrumental imperfections such as beam effects, reflections, and coupling artefacts further complicate this task by spreading the spectrally smooth foregrounds across the spectral and angular Fourier space. This makes both foreground avoidance and foreground subtraction more challenging. 

Recently proposed methods of dealing with the systematics involve either masking the systematic artefacts, filtering the affected Fourier modes out, or implementing instrumental changes such as altering cable lengths to shift the location of the reflection systematics. These methods have their own drawbacks, as discussed in Sect.~\ref{sec:intro}. Filtering, in particular, introduces spurious ringing and correlation features into observed statistics like the delay power spectrum, and can also result in signal loss \citep{Garsden2024}. The instrumental alterations, on the other hand, can be costly and can hinder array operations. They also do not alleviate the problems present in data that have already been observed prior to modifications. 

Bayesian inference approaches provide a unique solution to these problems by explicitly modelling the systematics instead of attempting to entirely filter them or ignore them, minimising signal loss and allowing for robust propagation of uncertainties associated with the systematics \citep{Murphy2024}. The forward modelling of systematics also avoids the introduction of additional artefacts in the delay power spectra since it provides us with continuous (unfiltered) realisations of the data model and its components to work with.

This work introduces a new module that can model multiplicative systematic effects as part of the \code{hydra-pspec} code \citep{Kennedy2023,Burba2024}, a Gibbs sampler that can already model the 21cm signal, foregrounds, and the 21cm delay power spectrum for interferometric visibilities. 

A Gibbs Sampler is a particularly useful tool in this instance, since it is capable of efficiently sampling large parameter spaces, potentially numbering in the millions of parameters. In effect, the full joint posterior distribution of all of the parameters can be sampled and analysed in a tractable manner.

As discussed in Sect.~\ref{sec:methods}, the foundation of this work lies in modifying and adding to the data model used in \cite{Burba2024}. We have treated the systematics contaminants as multiplicative gains on the sky visibilities. These are parameterised as the coefficients of modes in delay-fringe rate space, $\mathbf{b}_\textup{sys}$, which are then multiplied by a Fourier projection operator $\mathbf{H}$ for a specific and restricted set of modes. This parametrisation lets us model the systematics based specifically on the Fourier modes where the contaminants appear. This significantly reduces the number of parameters being sampled, improving computational efficiency, but requires a priori knowledge of where to place the systematics modes.

In Sect.~\ref{sec:results}, we demonstrated that the Gibbs Sampler, when implemented with the systematics model developed in this work, can accurately recover the EoR delay power spectrum from simulated data that contain different types of systematic contamination. When only one systematic artefact is present, the recovered EoR delay power spectrum is largely independent of the location of the systematics artefacts, although recovery of other parameters (namely, the foreground and systematics coefficients) is impacted by correlations between foregrounds and systematics, as discussed in Sect.~\ref{subsec:corr}. In the presence of multiple, distinct artefacts, the systematics coefficients themselves are recovered less accurately, while the recovered EoR delay power spectrum and the width of its posterior are not measurably affected. 

We also inspected the recovery of each of the individual components as shown in Fig.~\ref{fig:res_comp} and Fig.~\ref{fig:result_waterfalls}. The errors are broadly comparable to the noise realisation, as shown in Fig.~\ref{fig:res_comp}, which assures us that our solutions are reasonable, although some of the parameters are biased by several posterior standard deviations. This bias was ascribed to the foreground-systematics correlations and long correlation lengths of the Gibbs chains for two of the cases, and suggestions were made for how to reduce their impact. The combined systematics case (Sect.~\ref{subsec:res_combined}) reinforces this picture: parameters at foreground-adjacent delay positions again show the largest degradation, now across a broader set of Fourier modes simultaneously. The EoR delay power spectrum, though broadly accurate, also showed excessive scatter around the true model compared to the estimated uncertainties (Fig.~\ref{fig:dps}). This is likely due to underestimation of uncertainties due to omitting time correlations from our model, as discussed in \cite{Burba2024}.

Cases \RNum{1} and \RNum{3} exhibit the largest uncertainties because of a strong degeneracy between the foreground and systematic components at the level of individual Fourier modes: when the two overlap in delay or share fringe-rate structure, the sampler cannot easily attribute power to one component over the other. Case \RNum{2} is subject to a different degeneracy instead. Its short integrated autocorrelation times and large effective sample sizes indicate that the small offset present in its recovered systematics parameters does not arise from poor convergence of the Gibbs sampler. Rather, it reflects a degeneracy inherent to the multiplicative data model itself: the observations constrain only the product of the gain perturbation and the sky model, not the systematic parameters independently, so small changes in the recovered sky model can be compensated by corresponding shifts in the inferred systematic coefficients, while still preserving an accurate reconstruction of the visibilities and delay power spectrum.

The combined systematics case (Sect.~\ref{subsec:res_combined}) shows a related pattern and separates the two effects more clearly. The improvement in correlation time relative to the individual cases is not uniform across the twelve parameters: it is concentrated in those occupying the Case~\RNum{1} and Case~\RNum{3} modes, which share the same delays, and which speed up by factors of up to $11.9$, while the parameters at the well-separated Case~\RNum{2} delays are somewhat slower than they are in isolation. We stress that the sampler alternates between the sky and systematics conditionals in the combined case exactly as it does in the individual ones, and that the slowly-explored foreground-systematics direction identified in Sect.~\ref{subsec:corr} is still present and still traversed only by that alternation. We do not have a complete explanation for why the parameters sharing a delay should be the ones to speed up, and report the pattern as an empirical one. We note, however, that the systematics posteriors in the combined case are narrower than in the individual cases while being less accurately centred (Sect.~\ref{subsec:res_combined}), indicating that the combined run is overconfident about these parameters. This is the same tendency identified for the EoR delay power spectrum above.

The methods presented in this work consider the effect of cable reflections on a single interferometric baseline in order to isolate the behaviour of the systematic. In practice, however, HERA power spectra are formed by combining measurements from multiple baselines, including redundant baseline groups. Extending the present framework to the full array would therefore involve modelling the foreground and cable reflection response independently for each baseline prior to redundant averaging; since instrumental systematics are introduced at the visibility level, this represents a natural generalisation of the current methodology, one that would additionally capture the baseline-dependent extent of the foreground wedge and provide a more realistic assessment of cable reflection contamination in full-array observations.

The framework presented in this work is formulated for instrumental systematics that can be modelled as multiplicative modifications to the measured visibilities, with cable reflections serving as the primary application. Other multiplicative effects, such as gain errors or multiplicative signal loss terms, could likewise be incorporated through suitable models of their delay-frequency behaviour. More complex instrumental effects, however, including mutual coupling, beam-related systematics, residual low-level radio-frequency interference, and non-linear digitisation artefacts, are likely to require a more general, modified treatment.

A further, more practical limitation concerns the identification of the systematics itself, rather than its physical form: our method is dependent on the correct identification of the Fourier modes of the data affected by systematic contamination. Though this is absolutely known in this work by design, it may be more challenging in observational data, and an inaccurate identification of these modes will result in a systematics template that is less efficient or leaves some of the effects unmodelled. Encouragingly, the combined case in Sect.~\ref{subsec:res_combined} demonstrates that the method scales to at least 12 simultaneously-modelled modes without qualitative breakdown, though the practical challenge of correctly identifying all affected modes a priori naturally grows with their number. 

\section*{Acknowledgements}
This result is part of a project that has received funding from the European Research Council (ERC) under the European Union's Horizon 2020 research and innovation programme (Grant agreement No. 948764). MW was funded by a CITA National Fellowship.
We acknowledge use of the following software: 
{\tt matplotlib} \citep{matplotlib}, {\tt numpy} \citep{numpy}, and {\tt scipy} \citep{2020SciPy-NMeth}.

\section*{Data Availability}

The software described in this article is available from \url{https://github.com/HydraRadio/hydra-pspec} (branch: \code{systematic}).

\section*{Conflict of Interest}
The authors declare no conflict of interest. 

\balance


\bibliographystyle{mnras}
\bibliography{hydra-pspec-systematics} 


\bsp
\label{lastpage}
\end{document}